\documentclass[article]{JHEP3} 

\usepackage{epsfig,multicol}
\usepackage{amsmath}
\usepackage{graphicx}
\usepackage[latin1]{inputenc}
\usepackage{amssymb}
\usepackage{amsmath}
\usepackage{epsfig}
\newcommand\fverb{\setbox\pippobox=\hbox\bgroup\verb}
\newcommand\fverbdo{\egroup\medskip\noindent%
                      \fbox{\unhbox\pippobox}\ }

\newcommand\fverbit{\egroup\item[\fbox{\unhbox\pippobox}]}
\newbox\pippobox

\newcommand{\pythia}{{\sc Pythia}}
\newcommand{\herwig}{{\sc Herwig}}
\newcommand{\sherpa}{{\sc Sherpa}}
\newcommand{\mgme}{{\sc MadGraph/MadEvent}}
\newcommand{\alpgen}{{\sc Alpgen}}
\newcommand{\madgraph}{{\sc MadGraph}}
\newcommand{\madevent}{{\sc MadEvent}}
\newcommand{\be}{\begin{equation}}
\newcommand{\ee}{\end{equation}}
\newcommand{\ba}{\begin{eqnarray}}
\newcommand{\ea}{\end{eqnarray}}
\newcommand{\bt}{\begin{tabular}}
\newcommand{\et}{\end{tabular}}
\newcommand{\bfig}{\begin{figure}}
\newcommand{\efig}{\end{figure}}

\newcommand{\pt}{p_{\perp}}
\newcommand{\kt}{k_{\perp}}
\newcommand{\mt}{m_\perp}
\newcommand{\ET}{E_T}
\newcommand{\HT}{H_T}
\newcommand{\HTjet}{\HT^\text{jet}}
\newcommand{\HTrad}{\HT^\text{rad}}
\newcommand{\MET}{\not\!\!E_\perp}
\newcommand{\QME}{Q_\text{cut}^\text{ME}}
\newcommand{\Qmatch}{Q_\text{match}}
\newcommand{\Qhardest}{Q_\text{hardest}^\text{PS}}
\newcommand{\QMElow}{Q_\text{softest}^\text{ME}}

\title{QCD radiation in the production of heavy colored particles at the LHC}

\author{Johan Alwall\footnote{Research supported by the Swedish Research Council}\\
Stanford Linear Accelerator Center, \\
2575 Sand Hill Rd, Menlo Park, CA 94025, USA\\
E-mail: \email{alwall@slac.stanford.edu}
}

\author{Simon de Visscher, Fabio Maltoni \\
Centre for Particle Physics and Phenomenology (CP3) \\
Universit\'{e} Catholique de Louvain\\
Chemin du Cyclotron 2, B-1348 Louvain-la-Neuve, Belgium\\
E-mails: \email{simon.devisscher@uclouvain.be, fabio.maltoni@uclouvain.be}
}

\abstract{We present a study of effects of QCD radiation 
in the production of heavy colored states, employing inclusive
multi-jet samples obtained by matching matrix elements and parton
showers.  We discuss several examples showing that matched
samples are in general not only more accurate than a parton shower
alone, but also often indispensable to make reliable predictions of
beyond the Standard Model signals.
}


\begin{document}

\section{Introduction}

Many models for physics beyond the Standard Model, notably
Supersymmetry, Little Higgs models, various models for extra
dimensions and several versions of technicolor models, contain new
strongly interacting particles with masses below or near the TeV
scale. These particles, if they exist, will be copiously produced at
the LHC, and could provide the first and most important sign of new
physics. By analyzing their production and decay modes, we
might be able in only a few years to determine many of the new physics
properties, including large parts of the particle spectrum, spin
structure and the possible existence of new stable particles that can
act as dark matter candidates.

The final state signatures characteristic of events where heavy
colored particles are produced typically include a large energy
measured in the detector, missing transverse energy and several
hard jets coming from their decays. Such very spectacular events
can be efficiently simulated by Monte Carlo generators that implement
new physics models, including general-purpose generators like
\pythia~\cite{Sjostrand:2006za}, \herwig~\cite{Corcella:2000bw} and
\sherpa~\cite{Gleisberg:2003xi}, and matrix element generators such as
{\sc CalcHEP/CompHEP}~\cite{Pukhov:2004ca,Boos:2004kh},
\mgme~\cite{Alwall:2007st} and {\sc Whizard}~\cite{Kilian:2007gr}.
However, an additional difficulty in the simulation of particle
production at hadron colliders is due to the presence of abundant QCD
radiation, in particular initial-state radiation. This radiation,
which is enhanced in the production of heavy and strongly interacting
particles, can have important effects. It affects the event kinematics
by giving a transverse boost to the heavy particle system, and can
produce additional jets besides the jets originating from the decay of
the heavy particles, thus complicating the reconstruction and
indentification of the event.

This additional jet production has traditionally been simulated using
Parton Shower (PS) Monte Carlo programs such as \pythia\ and \herwig,
which describe parton radiation as successive parton emissions using
Markov chain techniques based on Sudakov form factors. This
description is formally correct only in the limit of soft and
collinear emissions, but has been shown to give a good description of
much data also relatively far away from this limit. However, for the
production of hard and widely separated QCD radiation jets, this
description breaks down due to the lack of subleading terms and
interference. For that case, it is necessary to use the full
tree-level amplitudes for the heavy particle production plus
additional hard partons.

The Matrix Element (ME) description diverges as partons become soft or
collinear, while the parton shower description breaks down when
partons become hard and widely separated. In order to describe both
these areas in phase space, the two approaches must be combined,
without double counting or gaps between different parton multiplicities. An
additional physical requirement is that such a procedure should give
smooth distributions, and interpolate between the parton shower
description in the soft and collinear limits and the matrix element
description in the limit of hard and widely separated partons. Several
algoritms have been proposed to achieve this, including the
CKKW~\cite{Catani:2001cc,Krauss:2002up},
L\"onnblad~\cite{Lonnblad:2001iq} and
Mangano~\cite{MLM,Mangano:2006rw} schemes. These different procedures
are in substantial agreement and give consistent results at hadron
colliders~\cite{Mrenna:2003if,Alwall:2007fs}.
For Standard Model processes, in particular the production of
jets and weak vector bosons, the matching schemes have been extensively 
used and compared to the available
data from the Tevatron~\cite{Krauss:2004bs}. In addition, inclusive 
top quark pair  production with matching has been compared to a 
next-to-leading order plus parton shower approach
(MC@NLO)~\cite{Frixione:2003ei} in Ref.~\cite{Mangano:2006rw}.

That jet matching is necessary for Standard Model backgrounds is
obvious, since the only way many Standard Model processes (such as
weak vector boson production) can emulate the production of new heavy
states charged under QCD is by taking into account high-$\pt$ radiated
jets. It is less clear, however, that jet matching should be important
in new physics production, for several reasons: The new particles are
expected to decay to high-$\pt$ jets making them fairly insensitive to
the effects of QCD radiation; furthermore the parton shower formalism
should work better the higher the mass is for the produced particles,
since then the region in which emissions can be considered
``collinear'' is increased. As we will demonstrate in this paper, this
is only partly true, and there are many scenarios where jet matching
turns out to be surprisingly important also in the production of heavy
new physics QCD states.

We will here use the production of squarks and gluinos within the
framework of the Minimal Supersymmetric Standard Model, MSSM. The
properties of initial state QCD radiation are however quite
insensitive to the precise type of particle produced, and mainly
depend on the mass scale and production mechanism of the new
particles. All our results here are therefore readily applicable to any
type of new physics producing colored heavy states decaying to jets.

The layout of the paper is as follows: In
Sec.~\ref{sec:technicalities} we discuss the details of the matching
schemes used, as well as how to solve the double counting problem that
arises due to the decay of on-shell states in multiparton final states. In
Sec.~\ref{sec:PSvsMEPS} we study the differences in QCD radiation between
matched and unmatched simulations. The matching is found to 
reduce the uncertainties of the parton shower  and hence to
increase the predictivity of the simulations. This increased
predictivity allows us, in Sec.~\ref{sec:anatomy}, to study the
effects of QCD radiation in the presence of jets from the decay of the
new states. We analyse in detail a few examples where jet matching
turns out to be crucial to get a qualitatively correct description of
the new physics signatures. In Sec.~\ref{sec:SPS1a} we take a look a
the impact of using matched simulations in a more realistic
experimental situation, where we include production of all relevant
supersymmetric particles as well as Standard Model backgrounds. 
We draw our conclusion in Sec.~\ref{sec:conclusions}.

\section{Technical challenges}
\label{sec:technicalities}

In this section we present the technical issues and the problems which
can arise while merging matrix elements involving heavy colored states
with a parton shower.  Our approaches and their implementations are
completely general and can therefore be applied equally well to the SM
as to any Beyond the SM (BSM) construction.  To be concrete we will focus 
on top pair production and on pair production of strongly interacting SUSY
particles, such as squarks and gluinos. In this study we have employed
the \mgme\ matrix element generator~\cite{Stelzer:1994ta,Maltoni:2002qb,Alwall:2007st}
interfaced to \pythia~4 \cite{Sjostrand:2006za} for parton showering
and hadronization. The MSSM implementation used was presented in
Refs.~\cite{Hagiwara:2005wg,Cho:2006sx,Alwall:2007st}.

\subsection{Matching schemes}
\label{sec:matching_schemes}

The goal of jet matching is to merge samples with different parton multiplicity
obtained via matrix elements, correctly accounting for showering
effects and avoiding double counting.  The matching algorithms used in
this study can be viewed as hybrids between the approaches currently
employed by \sherpa~\cite{Gleisberg:2003xi} and
\alpgen~\cite{Mangano:2002ea}. The phase space separation between the
different multijet processes is achieved using the $\kt$-measure
\cite{Catani:1993hr}. No analytic Sudakov reweighting of the events is
performed, but instead showered events are rejected if they are not
matched to the matrix element-level partons. This method allows the
use of the well-tuned showering and hadronization implementations of
\pythia\ while retaining the advantages of matrix element
production. The matching implementations in \mgme\ can be used both
for Standard Model and new physics processes.

The first matching scheme used in this study, the $\kt$-jet MLM
scheme, is the one used for \mgme\ in Ref.~\cite{Alwall:2007fs}. The
final-state partons in an event are clustered according to the
$\kt$-jet algorithm to find the ``equivalent parton shower history''
of the event. Here, the Feynman diagram information from \madgraph\ is
used to allow only clusterings that correspond to diagrams existing in
the generated matrix element. The smallest $\kt$ value is restricted to be
above some cutoff scale $\QME$. In order to closely mimic the behaviour
of the parton shower, the $\kt$ value for each clustering vertex
corresponding to a QCD emission is used as renormalization scale for
$\alpha_s$ in that vertex. As factorization scale, as well as
renormalization scale for the central hard $2\to1$ or $2\to2$ process,
the transverse mass $\mt^2 = \pt^2 + m^2$ of the particle(s)
produced in the central process is used.  This event is then passed to
\pythia\ for parton showering. After showering, but before
hadronization and decays, the final-state partons are clustered into
jets, again using the $\kt$ jet algorithm, with a cutoff scale
$\Qmatch > \QME$. These jets are then compared to the original partons
from the matrix element event. A jet is considered to be matched to
the closest parton if the jet measure
$\kt(\mathrm{parton},\mathrm{jet})$ is smaller than the cutoff
$\Qmatch$. The event is rejected unless each jet is matched to a
parton, except for the highest multiplicity sample, where extra jets
are allowed below the $\kt$ scale of the softest matrix element parton
in the event, $\QMElow$. This matching scheme can be used with both
the old (vituality-ordered) and the new ($\pt$-ordered) shower
implementations of \pythia.

In order to further study the systematics of the jet matching, a new
matching scheme has been implemented, which we call the ``shower $\kt$
scheme''. In this scheme, which is here used for the first time,
events are generated by \mgme\ as described above, including the
reweighting of $\alpha_s$. The event is then passed to \pythia\ and
showered using the $\pt$-ordered showers. For the $\pt$-ordered
showers, \pythia\ reports the scale of the first (hardest) emission in
the shower, $\Qhardest$. For events from lower-multiplicity
samples, the event is rejected if $\Qhardest$ is above the matching
scale $\Qmatch$, while events from the highest multiplicity sample are
rejected if $\Qhardest > \QMElow$, the scale of the softest matrix
element parton in the event. This matching scheme is simpler and yet
effectively mimics the workings of the $\kt$-jet MLM scheme. However,
it allows for the matching scale $\Qmatch$ to be set equal to the
matrix element cutoff scale $\QME$, and it more directly samples the
Sudakov form factor used in the shower. Furthermore, the treatment of
the highest multiplicity sample more closely mimics that used in the
CKKW matching scheme.

Since we here study heavy QCD particle production at hadron colliders,
we expect final-state radiation to be highly suppressed as compared to
initial-state radiation. The treatment of final-state radiation from
the (undecayed) produced particles is therefore of somewhat secondary
importance, and does not strongly affect any results of this study. We
have here chosen not to differentiate between shower particles
originating from final-state radiation and initial-state radiation in
the jet clustering in the $\kt$-jet MLM scheme, and correspondingly
for the ``shower $\kt$'' scheme, veto also events where the
final-state radiation reports an emission scale above the matching
scale.


\subsection{Double counting from resonant diagrams}

In the simulation of events where colored particles are produced that
can decay into one another by emitting partons, a further 
problem arises, also related to double counting.  
Consider, for example, the contributions to $\tilde g \tilde q$ + 1 jet coming from various
subprocesses. Among these, $gg \to \tilde g \tilde q\bar q$ displays a
peculiar behaviour. In Fig.~\ref{fig:one} we show three representative
diagrams out of a total of sixteen. Diagram (a) is a ``genuine''
correction to the $2 \to 2$ Born amplitude and it is correctly handled
by the matching procedure.  Diagrams (b) and (c), however, contain
possibly resonant gluino and squark propagators. When integrated over
the phase space these diagrams give rise either to $gg \to \tilde g
\tilde g$ with $\tilde g \to \tilde q \bar q $ or to
 $gg \to \tilde q \tilde q^*$ with $\tilde q^* \to \tilde g\bar q $,
 depending
on the mass hierarchy. These contributions, however, are already taken
into account by the Born level processes $gg \to \tilde g \tilde g$ or
$gg \to \tilde q \tilde q^*$ plus the corresponding decays and have
therefore to be properly subtracted to avoid double counting.  In
fact, this kind of issue is not specific to SUSY, but appears
every time a given final state can be reached through different decay
cascades.\footnote{A well-known example in the Standard Model is in
the calculation of the strong corrections to $tW$ production. In this
case the process $gg \to tWb$ contains a diagram with a resonant top
which overlaps with $t\bar t$ production~\cite{Tait:1999cf,Frixione:2008yi}} 
Several solutions have been proposed in the literature, with various degrees of approximations,
only a few of which can be employed in a Monte Carlo framework. For
example, subtractions at zero width, typically implemented in NLO
calculations, are not suitable to a Monte Carlo approach. In
Ref.~\cite{Plehn:2005cq} these diagrams were removed from the matrix
element by hand. Even though this procedure in principle violates
gauge invariance, it works very well in practice when the width of the
resonances is sufficiently small, $\Gamma/m \ll 1$. In order to deal
with this kind of problem in full generality we have implemented and
checked two independent types of solutions.

\FIGURE[t]{
\epsfig{file=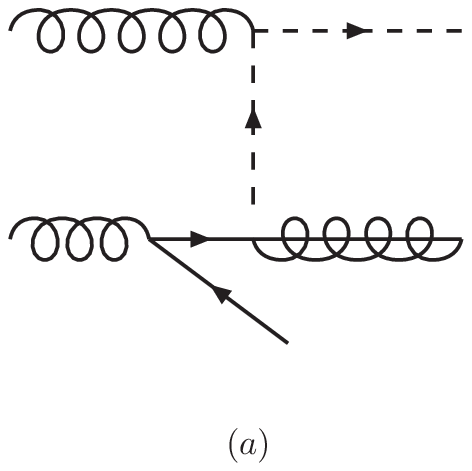 ,width=4.5cm}
\hspace*{.3cm}
\epsfig{file=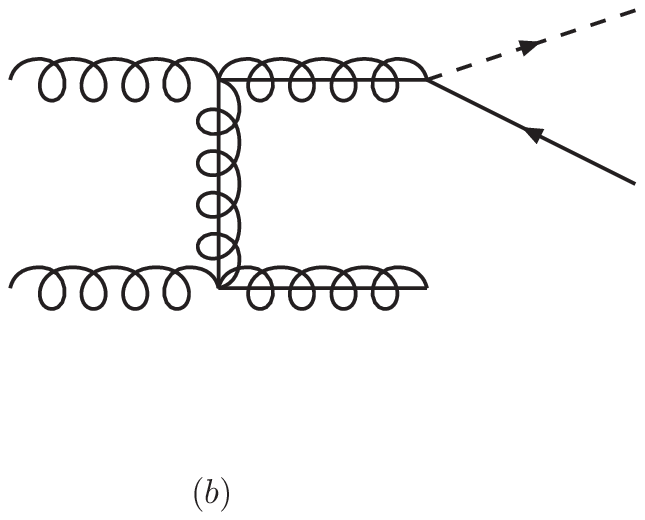 ,width=4.5cm}
\hspace*{.5cm}
\epsfig{file=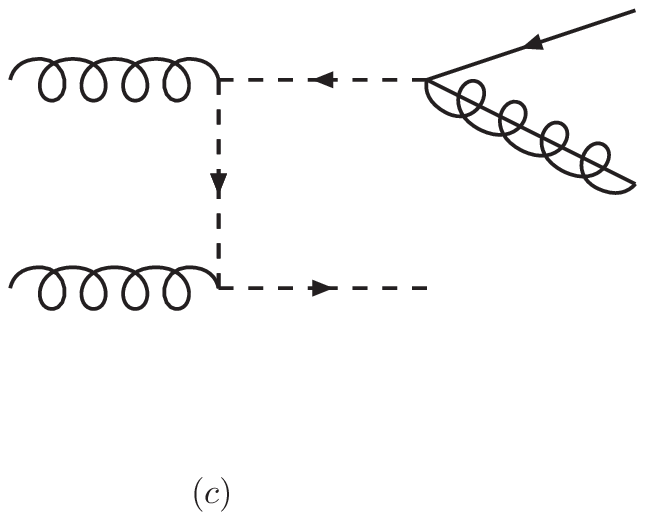 ,width=4.5cm}
\caption{Representative Feynman diagrams contributing to
$\tilde g \tilde q$+1 jet via the subprocess $gg \to \tilde g \tilde q q$. 
Diagram (a) is a ``genuine'' QCD correction to the Born level diagram
and correctly handled  by the matching procedure. 
Diagrams (b) and (c) belong to the same gauge invariant class 
as (a), however they contain a possibly resonant 
gluino (squark) contribution, relevant when
$m_{\tilde g}>m_{\tilde q}$ ($m_{\tilde g}<m_{\tilde q}$). 
These processes are already taken into account in the $\tilde g \tilde g$ and $\tilde q \tilde q^*$ channels, respectivetly, and have to be subtracted.
}
\label{fig:one}
}

In the first, resonant diagrams are removed as in 
Ref.~\cite{Plehn:2005cq}, but in an automatic way directly by MadGraph.
This approach has the virtue of being very simple, though it is not
gauge invariant and neglects the interference between the diagrams. 
The impact of these approximations therefore has to be carefully checked 
case by case.

The second solution is an hybrid between a proper
gauge-invariant subtraction and the diagram removal. It is based on a
general algorithm implemented in \madevent\ to record the
information of the presence of intermediate resonant propagators on an
event-by-event basis.  Such information is needed by the showering
program to ensure that the QCD evolution does not shift or even cancel
the Breit-Wigner peaks.  When only resonant or only non-resonant
diagrams are present, there is, in principle, no ambiguity and the
information could be directly passed to the events. When both resonant
and non-resonant diagrams are present such a unique assignment is not
possible, due to interference effects. However, the interference terms
are typically suppressed by $\Gamma/m$, and can therefore be neglected 
in the narrow width limit. In this case one can associate
the presence of a propagator to an event on a statistical basis, using
the relative size of the corresponding squared amplitudes.  The above
procedure exactly parallels that commonly employed to associate a
color flow to an event~\cite{Odagiri:1998ep}:  The calculation of the
matrix element is exact at all orders in the number of colors (as
the amplitude-squared includes all diagrams and their interference), 
however the color flow assignement is approximated by ignoring interference
terms which are proportional to $1/N_c^2$.
Such propagator/diagram mapping is possible in  \mgme\ due
to the fact that automatic integration over space space is 
already based on the amplitudes corresponding to single diagrams. 
The phase space integration is split into independent
channels, each corresponding to a single Feynman
diagram~\cite{Maltoni:2002qb}. Events generated in a resonant channel
are provided with the resonant particle information and also one of the
color flows consistent with the Feynman diagram itself (in so doing, we
avoid possible mismatches between the propagator structure and the
color flow, which would lead to an inconsistent shower evolution).
Having at our disposal the above procedure, it easy to get rid of the
resonant contributions, by simply dropping the events with resonant
propagators before passing them to the \pythia\ interface.
The results presented and discussed in the following are obtained using
the latter method. However, we have checked that the two methods are
consistent.

\section{QCD radiation in ME/PS merged samples}
\label{sec:PSvsMEPS}

QCD radiation in production of heavy states at hadron colliders is
expected to be mainly due to initial state radiation.%
\footnote{While this is strictly speaking a non gauge-invariant
statement, given that radiation is associated to color flows and not
to particles, it has become common jargon to talk about initial and
final state radiation. Such a separation is in fact meaningful only
for the collinearly enhanced radiation.}
The main reason for this is that the phase space for initial state
radiation is typically large and that collinear radiation from the
final state particles is suppressed by their large mass.
The main impact of QCD radiation on the production of heavy states is
therefore twofold: A transverse boost of the heavy particle pair, and
production of additional possibly hard jets radiated from the incoming
parton lines. The details of the initial state radiation are expected
to depend mostly on the mass of the produced particles, which 
sets the scale for the radiation, and on the type of initial
state partons (whether the production is from valence quarks,
quark-antiquark or gluon fusion).

In this section we compare the QCD radiation pattern as obtained from
a parton shower alone (\pythia) and from a matrix element plus parton
shower (\mgme\ + \pythia) approach, for some key processes with heavy
colored final states. To simplify the discussion and reduce the model
dependence, we study specific heavy final states which are left
undecayed, and give all distributions after parton showering but
without hadronization. We compare the systematic
uncertainties involved in parton showering by using the two different
showering implementations of \pythia, the ``old'' (virtuality-ordered)
and ``new'' ($\pt$-ordered) showers, with a range of shower parameters
similar to that in \cite{Plehn:2005cq}. Similarly, for the matching we
use the same shower implementations and parameter variations, and
furthermore use two different matching schemes, the $\kt$-jet MLM
scheme for virtuality-ordered showers and the ``shower $\kt$ scheme''
for $\pt$-ordered showers, as discussed in
Sec.~\ref{sec:matching_schemes}.

For most of the studies below, the shower parameter that is varied is
the starting scale of the shower. This is the most important parameter
to determine the hardness of radiation allowed in parton shower
emissions, and the default value has varied over the
years. Originally, the default scale was set to the factorization
scale $\mu_F^2$ of the process (typically $\mt^2 = \pt^2 + m^2$ of
the produced particles), inspired by the notion of (initial state)
parton showers being the deconvolution of the DGLAP equation for the
parton density functions. However, with the so-called Tune A
\cite{TuneA}, based on $W$ and $Z$ boson production at the Tevatron,
this was replaced by $4\mu_F^2$. Later (from v.~6.319), this upper bound was
increased, for certain types of processes, to $s_{pp}$, the center of
mass energy of the proton-(anti)proton collisions, in order to further
improve the description of extra jet radiation in hadron
collisions. Following the nomenclature of Ref.~\cite{Plehn:2005cq}, we
will in the following call showers with a starting scale corresponding
to $\mu_F^2$ ``wimpy showers'', and showers with the starting scale
$s_{pp}$ ``power showers''.

To study the impact of QCD scale choices, we also vary the
factorization and renormalization scales by a factor $1/2$ and 2, in
the matrix element as well as the parton showers.

We consistently simulate $X+$ jets for $X=\tilde g\tilde g,\tilde
q^{(*)}\tilde q^{(*)},\tilde g\tilde q^{(*)}$ and $t\bar t$, with up
to two extra partons from the matrix element simulations.

\subsection{Variation of matching parameters}

Before starting the comparisons between unmatched (showered-only) and matched
distributions, a series of ``sanity checks'' on 
the matched distributions were performed.
These include the requirement of  a smooth transition between the region of
phase space described by the shower (below the matching scale
$\Qmatch$) and the region described by matrix elements (above
$\Qmatch$), and stability of distributions as the matching scale is
varied.  Since both the matching implementations employed here rely on the
Durham $\kt$ measure to achieve the separation of the phase space, the
most revealing 
distributions to study their features are the differential jet rates defined 
according to the same measure. In particular, in the $\kt$-jet MLM scheme 
there is at parton level a sharp division in the jet rates between the
shower and matrix element regions, making
it very easy to see to which extent the transitions are smooth. For
the shower $\kt$ scheme, as well as the cone jet MLM scheme
implemented in \alpgen\ and the CKKW scheme implemented in \sherpa, the
separation is less sharp, but the differential jet rates still tend to be
the best variables to study the transition between parton showers and
matrix elements.

Our guidelines for the choice of the scales $\QME$ and $\Qmatch$ are
based on smoothness of distributions across the matching transition as
well as efficiency. The higher the scale can be chosen, the higher
will the proportion of lower-multiplicity events be in the combined
sample. Since the computational effort is heavier for higher
multiplicities, it is desirable to choose the matching scale as high
as possible. Therefore for each choice of particle type, particle
mass, and shower, the matching scale is chosen to be close to the
highest scale that still gives a stable matched cross section and
smooth differential distributions, in particular differential jet rate
distributions. As it turns out, the ``new'', $\pt$-ordered \pythia\
showers allow significantly higher choices of the matching scale than
the ``old'', virtuality-ordered showers. The reason for this is that
they give significantly harder emissions than the old showers, and
therefore give distributions more similar to the matrix element
distributions.

In the following, our default scale choices are:
\begin{itemize}
\item For pair production of SUSY particles with heavy mass (600 GeV
and above):
For virtuality-ordered showers, $\QME = 40$ GeV and $\Qmatch=60$ GeV,
for $\pt$-ordered showers $\QME$ and $\Qmatch= 100$ GeV.
\item For Standard Model $t\bar t$ production or light gluino par production: For virtuality-ordered
showers, $\QME = 20$ GeV and $\Qmatch=30$ GeV, for $\pt$-ordered
showers $\QME$ and $\Qmatch= 100$ GeV.
\end{itemize}

\FIGURE[t]{
\epsfig{file=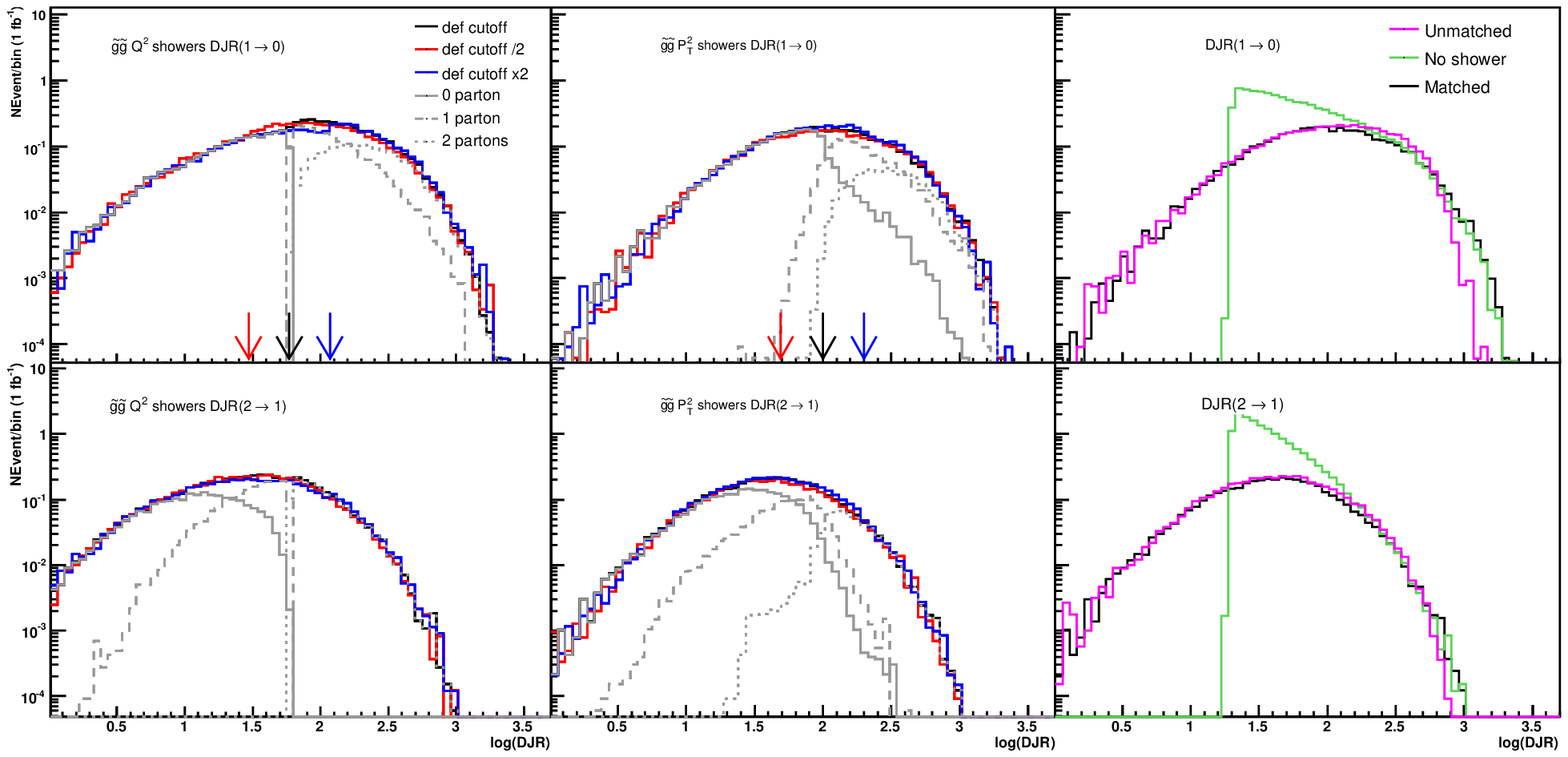 ,width=15cm}
\vspace*{.5cm}
\epsfig{file=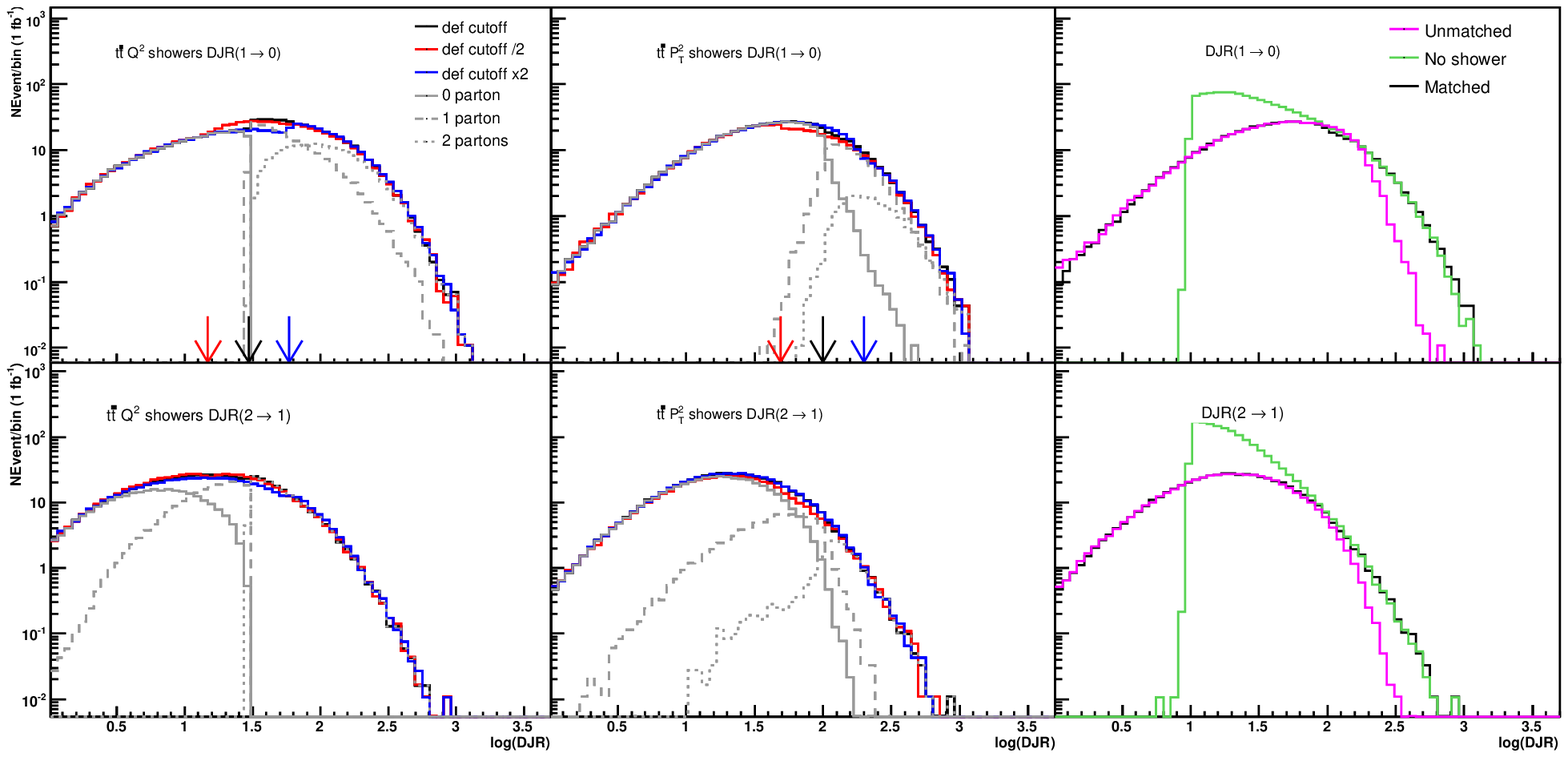 ,width=15cm}
\caption{Differential jet rates for $0\to1$ and $1\to2$
jets from QCD radiation for $\tilde g\tilde g$ and $t\bar t$
production at the LHC. The first two columns show the distributions
for the two types of \pythia\ showers, virtuality-ordered and
$\pt$-ordered, for three different choices for the matching scale. For
the default choice, also the contributions from the separate
multiplicity samples are shown. The colored arrow show the value of
the three $\Qmatch$ used for each kind of shower. The third column
shows how the matched curve interpolates between the pure parton
shower curve and the pure matrix element curve without parton
showering.}
\label{fig:two}
}

In Fig.~\ref{fig:two} we show the effects of varying the matching
scale by a factor $1/2$ and $2$ from the default values, for 607 GeV
gluino pair production and Standard Model top pair production, for the
new and old \pythia\ showers. We have here used the default starting
scales for the respective \pythia\ showers, corresponding to the
factorization scale $\mu_F$ for the $\pt$-ordered showers and $2\mu_F$
for the virtuality-ordered showers. 

In the right-hand column, the curve for unmatched \pythia\ showers with
default parameters are shown, together with the pure matrix element
prediction without any parton showering or matching and the matched
curve. We see, as expected, that the matched curve smoothly
interpolates between the unmatched \pythia\ curve below the matching
scale, and the matrix element prediction for large scales. 

We also see how the variation in the matched curves is small as the
matching scale is varied, and the curves above and below the matching
scale variation limits are quite stable.

The difference between the two matching schemes is visible in the
plots, in the different behaviour of the parton multiplicity sample
contributions. The lefthand column for each particle type shows the the
$\kt$-jet MLM matching scheme, with the contributions from the
different parton multiplicity samples in grey. The matching scale
cutoff is, in this scheme, done in the same variable that is plotted,
the differential jet rate, and there is therefore a sharp cutoff
between the 0- and 1-parton samples in DJR($0\to1$), and between the
1- and 2-parton samples in DJR($1\to2$), so that below the cutoff only
the lower-multiplicity samples contribute and above the cutoff only
the higher-multiplicity samples. In the middle column, the ``shower
$\kt$'' scheme is used (with the $\pt$-ordered \pythia\ showers). This
scheme cuts on the first emission of the parton shower rather than on
the combined radiation of the whole shower, giving some smearing
across the matching scale. This scheme therefore allows to use the
same cut at matrix element level and matching level. The distributions
for the $\pt$-ordered showers have been double-checked using the $\kt$-jet MLM
matching method, with excellent agreement.

Interesting to notice is the differences in curve shapes depending on
the choice of shower type. Below the matching scale, the shape of the
curve is given completely by the shower, in particular for the $0\to1$
jet rate. Above the matching scale, however, the shape is mainly given
by the matrix element. It is easy to see the reason for the different
choices of matching scales for the different showers -- the
$\pt$-ordered shower gives significantly harder distributions than the
virtuality-ordered shower, and is more similar to the matrix element
curve, hence allowing a higher matching scale.

\subsection{Parameter dependence in matched and unmatched generation}

\begin{center}
\FIGURE[t]{
\epsfig{file=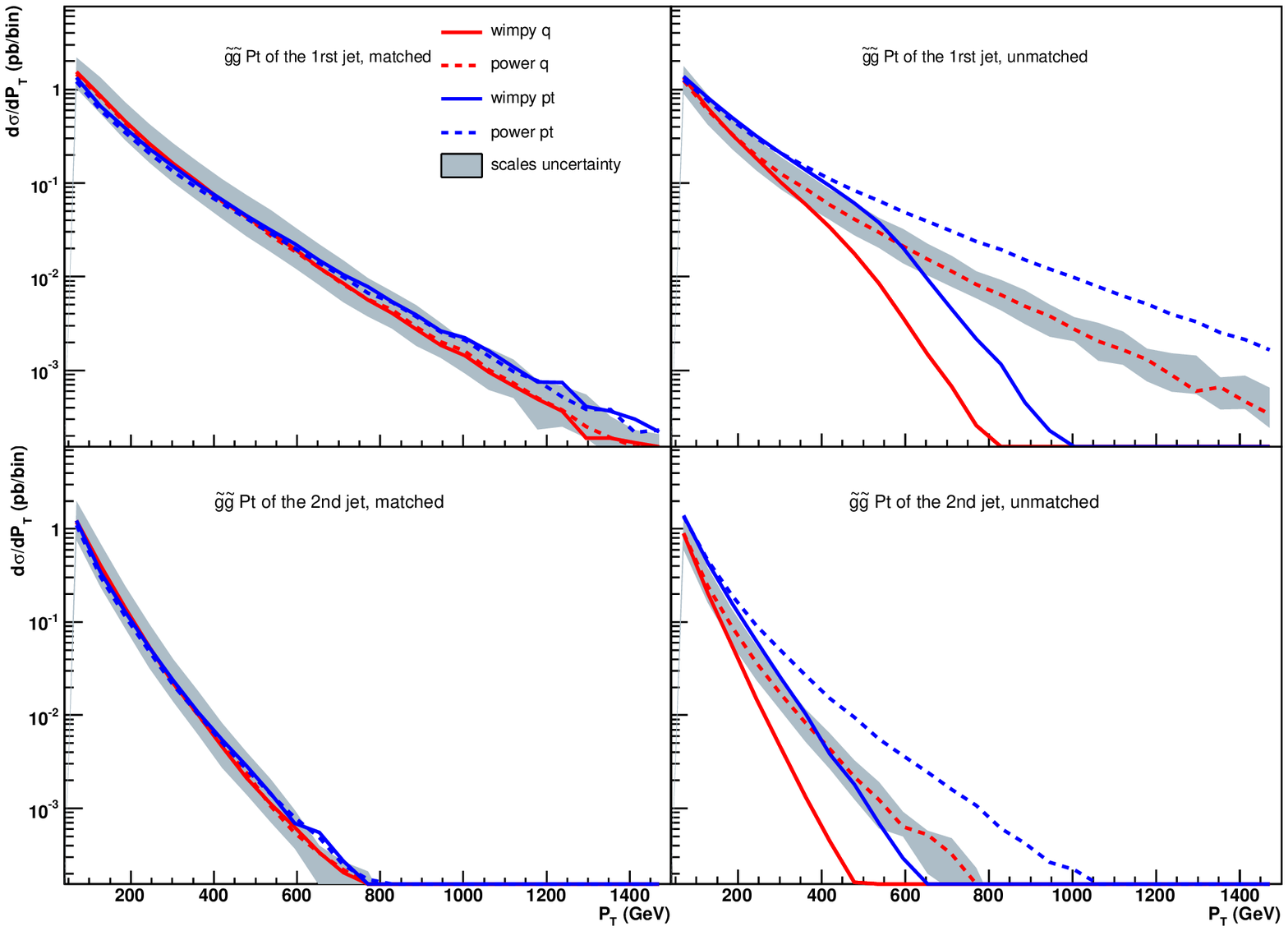 ,width=10cm}
\hspace*{.45cm}
\epsfig{file=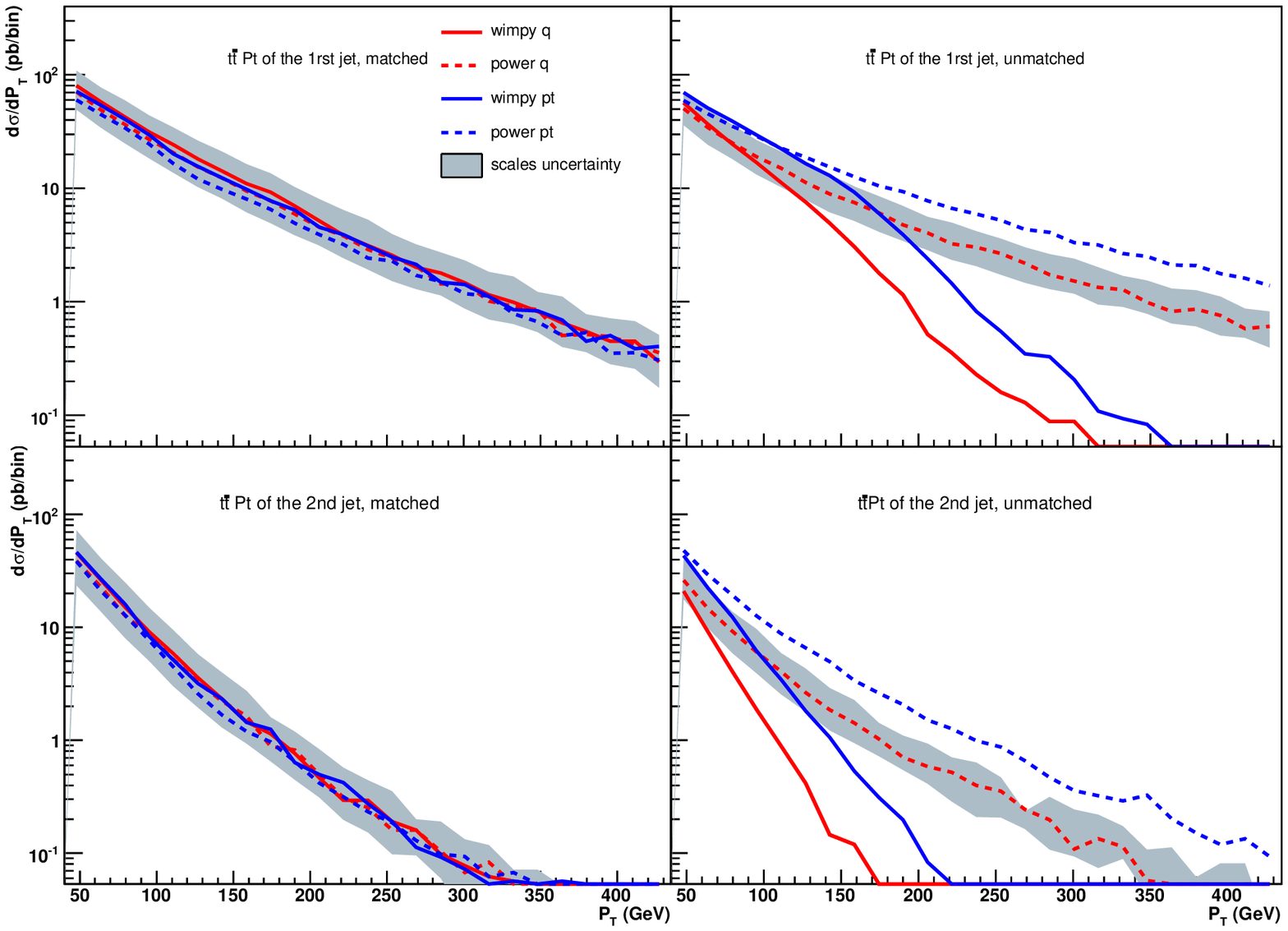 ,width=10cm}
\caption{$\pt$ spectrum for the first and second hardest
radiated jet in (a) $\tilde g \tilde g$ and (b) $t\bar t$ events. The
right column in each group of plots shows the spread of \pythia\ predictions
with different choices of the shower evolution variable
(virtuality- and $\pt$-ordered) and starting scale for the evolution
(labeled as ``wimpy'' and ``power'' showers respectively). The left column
presents the results obtained after matching in the same
four shower scenarios. The grey band shows the systematic uncertainty
associated with a variation of a factor of two of the renormalization
and factorization scales with respect to their central values. The
different curves have a normalization corresponding to their
cross section. The gluino mass is here 607 GeV, while the top
mass is set to 174 GeV.}
\label{fig:three}
}
\end{center}

One of the advantages of the parton shower formalism, and also one
of the arguments for using it, is that parton shower Monte Carlo 
generators have several parameters on which  
the behaviour of the shower depends, that can be tuned to the data. 
While this is certainly an advantage in general, it also
means that the parton shower lacks predictability at least for
some observables or areas of the phase space. It is not
always clear that a tune done for one type of initial state will
be applicable to other initial states, or that a tune done for a
particular mass of a pair-produced particle will be applicable for
other masses.

We first compare the matched and unmatched spectra of the first and
second jet $\pt$ in Standard Model $t\bar t$ production events and SUSY
$\tilde g\tilde g$ production with $m_{\tilde g}=607$ GeV, for
different shower parameters, Fig.~\ref{fig:three}. Here, we look at
only jets coming from QCD radiation, leaving the heavy particles
undecayed, in order to specifically study the differences in treatment
of radiation only. The different curves correspond to the different
shower parameters settings: the ``old'', virtuality-ordered shower and
the ``new'', $\pt$-ordered implementation, each with two different
choices for the starting scale of the shower, $\mu_F$ (``wimpy
shower'') and $s_{pp}$ (``power shower''). In order to study the
intrinsic QCD uncertainty on the predictions, we also vary the
factorization and renormalization scales for one of the parameter
settings by a factor 2 up and down. Note that we here vary all scales
together, both for the central process and for the parton shower/QCD
radiation in the matching.

Several interesting features can be noted from Fig.~\ref{fig:three}:

\begin{itemize}
\item The spread in predictions for the parton shower is very large
and strongly affects the shapes of the distributions. This uncertainty
due to shower parameters is almost completely removed when
matching is applied.
\item The region where the shower predictions start to diverge, and
the rate of this divergence, is strongly correlated with the mass of
the produced particles. This correlation is due to the choice of
starting scale for the ``wimpy'' showers as the factorization scale,
which is close to the mass of the produced particle.
\item The ``power'' shower curves consistently overshoot the matched curves,
and hence give too hard predictions, while the ``wimpy'' showers
give too soft distributions.
\item The uncertainty due to scale variations is considerable, but
mainly affects the normalization and only to a small degree the
shape of the curves.
\end{itemize}

The present \pythia\ default for the
virtuality-ordered showers is close to the curve for the ``wimpy''
$\pt$-ordered showers, which is also the default for the $\pt$-ordered showers.

\subsection{Impact of different inital states}

\FIGURE[t]{
\epsfig{file=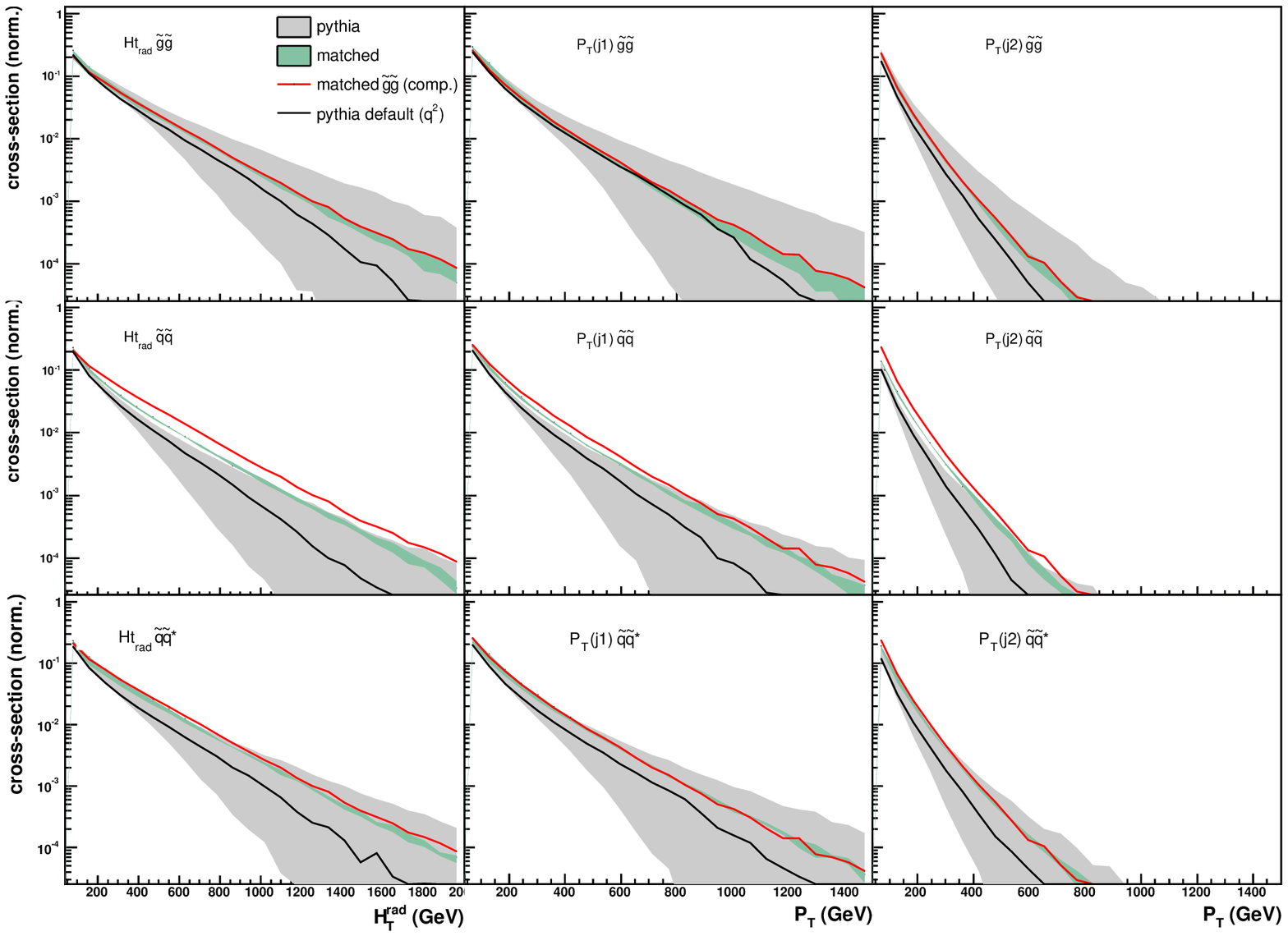 ,width=15cm}
\caption{$\HTrad$ and $\pt^\text{jet}$ distributions for radiated jets in
 $\tilde g \tilde g$, $\tilde q \tilde q$ and $\tilde q \tilde q^*$
 production for $m_{\tilde g, \tilde q} = 607$ GeV. The grey band
 shows the spread of unmatched \pythia\ predictions with varying
 shower parameters, while the green band shows the corresponding
 matched predictions. The full black curve represents the default
 unmatched \pythia\ $Q^2$-ordered shower, while the full red curve is
 the default matched curve for $\tilde g\tilde g$ production, included
 in the other plots as a guide for the eye. Each curve has
 normalization relative to the Born cross section, which in turn is
 normalized to unity in order to allow comparison between the different
 samples.}
\label{fig:four}
}

The impact of different initial states is shown in
Fig.~\ref{fig:four}, where we compare radiation for pair production of
gluinos, up-type squarks and up-type squark-antisquark. For ease of
comparison, the supersymmetric particle masses have all been set to
607 GeV. These different SUSY particles are produced by different
initial states: gluino production is mainly through gluon fusion,
squark-squark production from valence squarks exchanging a
t-channel gluino, and squark-antisquark production from a combination
of gluon fusion and quark-antiquark with t-channel gluio exchange. 
Besides the $\pt$ of the two hardest radiated jets, we also
show the total $\HT$ of radiated jets, defined as
\begin{equation}
\HTrad = \sum |{\pt}_i|  \\
\end{equation}
where the sum is taken over radiated jets with $\pt > 40$ GeV, as a
measure of the total QCD radiation activity.

An interesting observation from Fig.~\ref{fig:four} is that there is
no choice of shower parameters that describe all three types of
particle production. For squark pair production in particular, the
shower always undershoots the matched prediction, except for the
hardest shower choice ($\pt$-ordered power shower), which is much too
hard in the other productions. Indeed the matched curves are much more
similar between the different productions than the shower curves, in
particular in the tails (as illustrated by the inserted matched $\tilde
g\tilde g$ comparison curve). This indicates that the shower retains a too
strong ``memory'' of the initial state, while the matrix element
displays a larger independence from the initial state.

It is also interesting to note that the default virtuality-ordered
\pythia\ shower undershoots all curves, but least so for gluino
production and most notably for squark-squark production, where it
undershoots the matched curve by at least 50\% starting already
below a $\pt$ of 200 GeV.

\subsection{Produced particle mass dependence}

\FIGURE[t]{
\epsfig{file=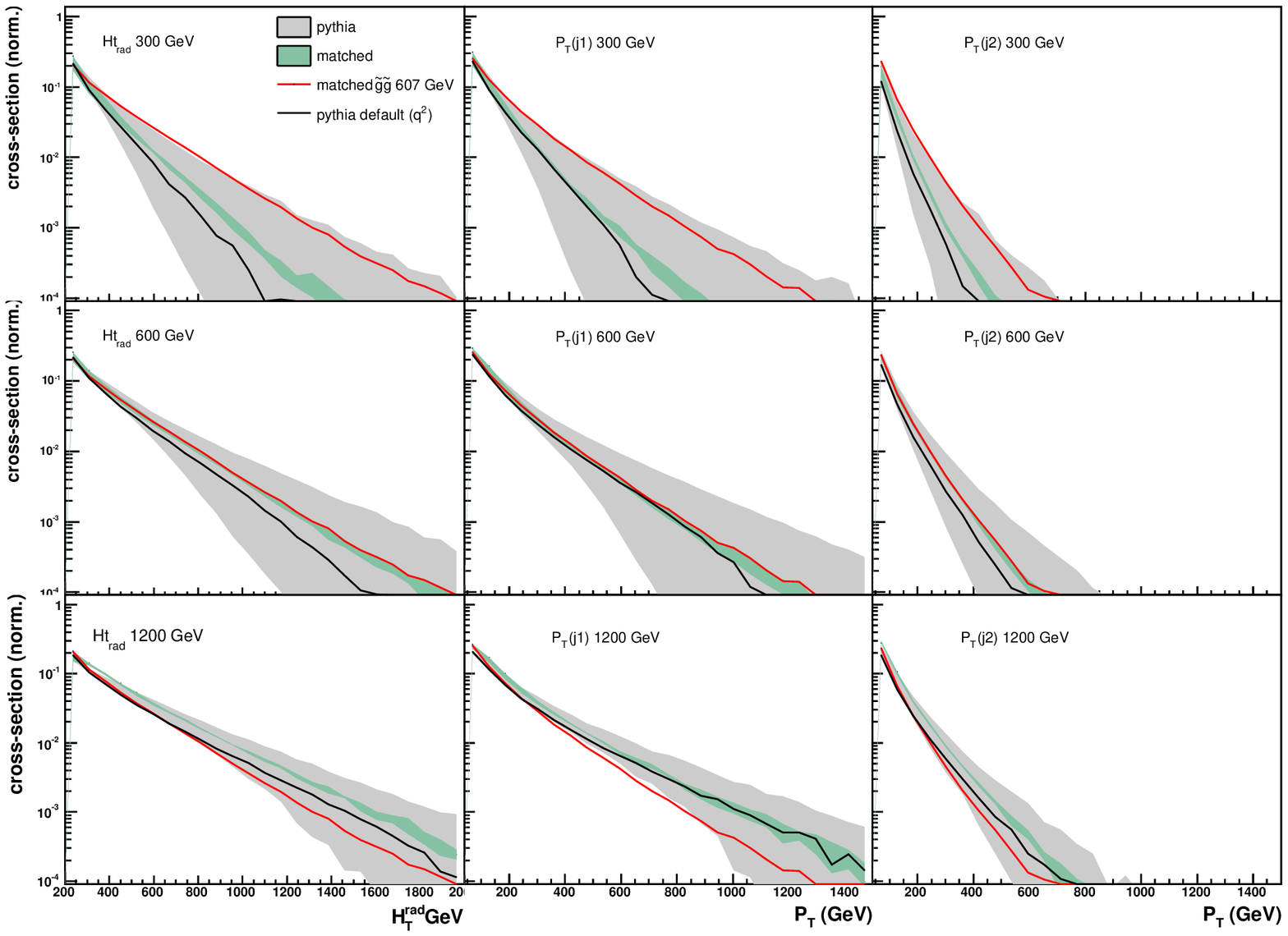 ,width=15cm}
\caption{$\HTrad$ and $\pt^\text{jet}$ distributions for radiated jets in
$\tilde g \tilde g$ production for $m_{\tilde g} = 300,600,1200$
GeV. The grey band shows the spread of unmatched \pythia\ predictions
with varying shower parameters, while the green band shows the
corresponding matched predictions. The full black curve represents the
default unmatched \pythia\ $Q^2$-ordered shower, while the full red
curve is the default matched curve for $m_{\tilde g} = 600$ GeV,
included in the other plots as a guide for the eye. Each curve has
normalization relative to the Born cross section, which in turn is
normalized to unity in order to allow comparison between the different
samples.}
\label{fig:five}
}

Determination of the absolute masses of produced particles at the LHC
is suprisingly non-trivial if the produced particles decay to an
invisible stable massive particle such as a WIMP. This difficulty is
due to the fact that only the kinematics of decays are due to mass
differences only, and are relatively independent of the absolute
masses involved.  Many methods have been devised to extract this
information, including the use of different transverse mass quantities
(see, e.g., \cite{Barr:2007hy,Cho:2007dh}) and combined information on
kinematics, cross sections and branching ratios
\cite{Lafaye:2004cn,Lester:2005je}, although
especially the latter tend to be quite model dependent. Another
possiblility would be to take advantage of the spectrum of QCD
radiation, since this is directly sensitive to the absolute mass scale
rather than mass differences. Studies have been done on the effect of
QCD radiation in the decays of squarks and gluinos
\cite{Miller:2005zp,Horsky:2008yi}. In a hadron collider however, the
majority of visible radiation originates from the initial state.

As can be seen from Fig.~\ref{fig:five}, the predictions for the
shapes of the spectra of radiated jets are quite precise once matching
is taken into account, while it is clear that this type of study
cannot be done with \pythia\  radiation only.

For the lightest SUSY scenario (300 GeV gluinos), the distribution of
the leading radiation jet is quite distinctive from the heavier
scenarios, while the difference between the 600 GeV and 1200 GeV
scenarios is small. Even barring the difficulties to distinguish
radiation jets from jets from decays, it would be very difficult to
differentiate between different high-mass scenarios using the
distribution of radiated jets. One can however imagine scenarios where
this could be a useful crosscheck, e.g.\ in the case where the mass
splitting between the produced QCD particle and the stable WIMP is
relatively small (such a scenario will be revisited in
Sec.~\ref{sec:degenerate} below). More work on this would clearly be
needed, but since this is beyond the scope for this paper, we
choose to save it for a future publication.

\section{Anatomy of $\MET$ + multi-jet final states}
\label{sec:anatomy}


\begin{table}
\begin{center}
\begin{tabular}{| l r |c|c|c|c|}
\hline
Scenarios  && I & II & III & IV  \\
\hline
 Masses (GeV) &
 \parbox{5mm}{$\tilde g$\\$\tilde q$\\$\chi_1^0$} &
  \parbox{10mm}{600\\550\\100} &
   \parbox{10mm}{600 \\ heavy\\100} &
    \parbox{10mm}{heavy\\550\\100} &
    \parbox{10mm}{600\\ heavy\\500} \\
\hline
\end{tabular}
\end{center}
\caption{\label{tab:benchmarks}
Benchmark scenarios employed in this work as modifications
of SPS1a~\cite{Allanach:2002nj}. 
We always assume the squarks decaying 100\% into quark+lightest neutralino.}
\end{table}

While the previous section dealt with jets from QCD radiation only, in
models with new heavy QCD states produced at the LHC we expect jets
also from their decays. It is therefore necessary to study the
impact of jet matching in the context where we include also such
jets. One could expect that jets from QCD radiation should be
relatively unimportant compared to the hard decay jets from heavy
particles, but as we will show, there are many situations when this is
not true. Indeed, as it turns out, there are surprisingly many
scenarios where radiation jets can significantly alter the analysis or
interpretation of data.

In order to clarify the discussion, and keep our results conservative,
in this section we will use a set of simplified supersymmetric
benchmark scenarios, summarized in Table~\ref{tab:benchmarks}. In all
the scenarios we assume all light-flavour squarks to have the same
masses and that they all decay directly to the LSP, i.e.\ we ignore the
existence of intermediate weak states. Introduction of cascade decays
will have as main effect that jets from decays get softer, while the
jets from QCD radiation are not affected, and will hence mainly
further accentuate our results.

For ease of comparison between the scenarios, we have chosen to use
the same masses, around 600 GeV, for the active heavy QCD states in
all scenarios. Scenario I has a SUSY QCD spectrum similar to the SPS
point 1a~\cite{Allanach:2002nj}, 
with a gluino at 607 GeV which decays to squarks at 560 GeV,
while the LSP is at 100 GeV. In scenario 2, the gluino has a mass of
607 GeV but all squarks are heavy, so that the gluino decays through
offshell squarks to two quarks and the LSP. Scenario III has squarks at
560 GeV and the gluino too heavy to be produced at the LHC, and
finally scenario 4 has a gluino at 607 GeV decaying through offshell
heavy squarks, but the LSP mass is 500 GeV, only 100 GeV lighter than
the gluino. These scenarios will act as ``cartoons'' to illustrate
different effects of QCD radiation in the production of new heavy QCD
states.

All plots and results in this section are generated using matched
samples with the default \pythia\ parameter choices for
virtuality-ordered showers unless otherwise stated.

\subsection{$\HTjet$ variables in gluino production}

\FIGURE[t]{
\epsfig{file=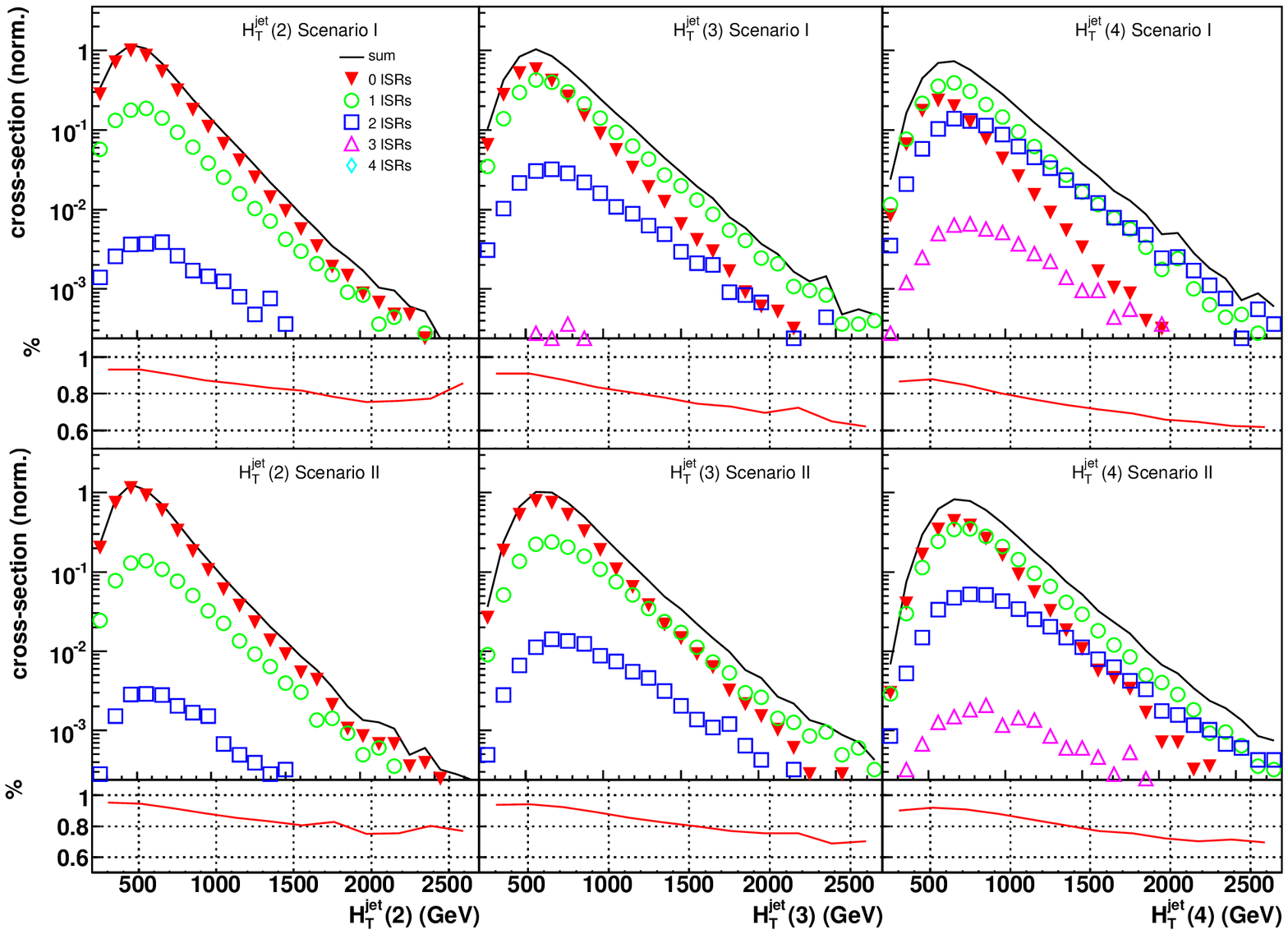,width=15cm}
\caption{$\HTjet(n)$ for $n=3,4,5$
in $\tilde g\tilde g$ production.  The different point markers show how many of
the jets entering in the definition of $\HTjet(n)$ come from QCD
radiation. Upper row: Scenario I, lower row: Scenario II. Below each main plot, the red curve indicates the percentage of $H_T$ coming from the decays. Jets are
defined using the SISCone algorithm with a $\pt^\text{min}$ of 40 GeV
and a radius of 0.5.}
\label{fig:six}
}

It is something of common lore that, in order to study squark pair
production one should select 2-jet observables, for associated gluino-squark production at 3-jet observables and for gluino pair
production at 4-jet observables. While it is obviously
true that gluinos must decay to two quarks and a color singlet
(barring the exotic possibility that the dominant decay is the
loop-mediated twobody decay to a gluon and a color singlet), it should
be kept in mind that the visibility of these jets depends strongly on
the mass hierarchy of QCD states. 
In Fig.~\ref{fig:six} we show the $\HTjet(n)$, defined as
\begin{equation}
\HTjet(n) \equiv \sum_{i=1}^n |{\pt}_i^\text{jet}|
\label{eq:htjet}
\end{equation}
for $n=2,3,4$, for the scenarios 1 and 2. The sum in
eq.~(\ref{eq:htjet}) is taken over jets defined using the
SISCone~\cite{Salam:2007xv} algorithm with a radius of 0.5 and
$\pt>40$ GeV. We also show the composition of $\HTjet(n)$ in terms of
jets from the gluino decay and radiated jets (``ISR''), as well as
show the average fraction of the $\HT$ coming from the decay. For
scenario 2, where the gluinos decay through off-shell squarks to two
quarks of similar energies and an LSP, the majority of events in the
peak of $\HTjet(4)$ include only jets from the decay, while the tail
of the distribution is dominated by 2-3 jets from the decay and one
jet from radiation. For scenario 1 however, where the 600 GeV gluino
decays into a fairly soft jet (with an energy around 50 GeV) and a
squark which in turn decays to a hard jet and an LSP, the
distributions are quite different. Here, $\HTjet(4)$ is dominated by
events where at least one of the jets come from radiation, and we need
to go down to $\HTjet(2)$ to be dominated by events with only decay
jets across the whole $\HT$ range. The energy fraction of $\HTjet(4)$
coming from decay is still fairly high, even in the high-$\HT$ tail,
since most of the transverse energy comes from the squark decay jets.

The immediate interpretation of this result is that in a scenario with
a small mass splitting between gluinos and squarks, gluino production
might be difficult to distinguish from squark production with
additional QCD radiation. In Section~\ref{sec:falsegluino} we will
study this in greater detail.

\subsection{Jet multiplicities for different scenarios}\label{jetnumsec}

A question that the proper matching of jets is particularly apt to
answer, is to specify the number of jets typically present in
production of different particles. We here present a table with the
jet multiplicities, for matched and unmatched (Pythia virtuality-ordered
default) production, for the scenarios studied. In order to make
the table as useful as possible, we have used jet cuts close to what
is used in many preparatory analyses for squark and gluino searches:
$\pt^{\mathrm{jet}_1}>180$ GeV, $\pt^{\mathrm{jet}_2}>110$ GeV,
$\pt^{\mathrm{jet}_j}>50$ GeV for $j>2$ and $|\eta^\mathrm{jet}|<3$
for all jets. We require all events to have at least two jets. The jet
multiplicities are 
exclusive, and so add up to 100\% of the events passing the 2-jet
cut. The jet algorithm used is SISCone with a radius of 0.5 and $\pt>40$ GeV.

\TABLE[ht]{
\begin{tabular}{|l||c|c|c|c|c|c|c|c|c|c||c|c|}
\hline
Process&\multicolumn{2}{c|}{$N=2$} & \multicolumn{2}{c|}{$N=3$}
& \multicolumn{2}{c|}{$N=4$} & \multicolumn{2}{c|}{$N=5$} &
\multicolumn{2}{c||}{$N\geqslant 6$}&\multicolumn{2}{c|}{Signal eff.}\\
\hline
&M&U&M&U&M&U&M&U&M&U&M&U\\
\hline
\hline
$\tilde{g}\tilde{g}$ sc.I&15.7&27.1&30.0&33.9&24.4&21.0&13.0&8.9&16.7&8.9&43.7&40.4\\
$\tilde{g}\tilde{q}$ sc.I&35.2&39.4&32.5&33.8&17.5&16.0&7.0&5.2&7.64&5.2&31.9&28.3\\
$\tilde{q}\tilde{q}$ sc.I&40.2&48.1&33.4&32.0&15.9&12.9&4.9&3.8&5.4&3.4&16.9&16.0\\
$\tilde{g}\tilde{g}$ sc.II&4.4&4.7&19.5&22.1&27.1&29.2&18.0&17.6&31.1&26.1&43.8&40.1\\
$\tilde{g}\tilde{g}$ sc.IV&21.5&28.4&32.6&37.0&23.9&21.0&11.4&7.8&10.4&5.8&4.7&3.0\\
\hline
\end{tabular}
\caption{Contribution of events with $N$ jets for matched and unmatched
processes: $\tilde g \tilde g$, $\tilde{g}\tilde{q}$ and
$\tilde{q}\tilde{q}$ in Scenario I (for squark production this is
very similar to scenario III), and $\tilde g\tilde g$ in Scenario II and
IV. All numbers are in percent. ``Signal efficiency'' shows the
percentage of events that pass the 2-jet cut. The jet cuts are
described in the text.\label{tab:perc}}
}

The difference between matched and unmatched generation (as seen in
Fig.~\ref{fig:four}) is twofold: First matching tends to increase the
transverse boost of the produced pair, and hence the $\pt$ of the
softest jets from the decays. Second, the ISR jets get harder, and
more easily get above the threshold of 50 GeV. The matched production
therefore in general populates higher jet number bins the unmatched ones.

The first three lines in Table~\ref{tab:perc} represent Scenario I,
which is very similar to the benchmark point SPS1a. In this scenario,
the gluinos decay as $\tilde g \rightarrow \tilde q q
\rightarrow\tilde q q \chi^0$ (with $m_{\chi^0}\sim$100 GeV), so there
are typically two hard and two soft jets from the gluino decay. Here
the effect of the matching is large, due to the increase in $\pt$ for
the soft jets as well as increased hardness of the ISR jets. For
scenario II on the other hand, where gluinos decay in three body decay
into two jets and a light neutralino, the sensitivity to the matching
is much lower, since there, typically at least four reasonably hard
jets are present from the decays. It is only in this and similar cases
that the statement that gluino pair production generally corresponds
to four hard jets in the event is true. In the last row in the table,
the produced gluinos decay to a near-degenerate LSP, meaning that most
hard jets are due to QCD radiation. Here, a large recoil against
initial state jets is needed in order to even pass the 2-jet cut,
hence the very low signal efficiency.

For squark-squark and gluino-squark production for Scenario I (row 2
and 3 in Table~\ref{tab:perc}, there are only two hard partons from
the decays (and one additional soft parton from the gluino decay) and
additional QCD radiation. Again the main consequence of the matching is
to increase the mean number of jets. It is interesting to note that
in this particular case, the addition of matching to the generation of
squark pair production gives very similar numbers as the unmatched
$\tilde g\tilde q$ associated production (although the selection
efficiency is different), indicating that the matching has a similar
impact as the addition of one extra jet with $\pt \sim 50$ GeV.

We now look closer at two examples where an analysis based only on a
parton shower approach might lead to the wrong conclusions.

\subsection{Example 1: False gluino evidence}
\label{sec:falsegluino}

We here consider the case given in Scenario III, where the only
observable SUSY particles are 560 GeV squarks, while the gluinos are
too heavy to be produced. If such a scenario is simulated using only
parton showers, the result will typically be a deficit in the number
of events with multiple jets. This is illustrated in
Fig.~\ref{fig:seven}, where we show $\HTjet(2)$, $\HTjet(3)$ and
$\HTjet(4)$ (defined in eq.~(\ref{eq:htjet})) for matched and
unmatched simulation of $\tilde q\tilde q^{(*)}$ production, using the
\pythia\  virtuality-ordered shower with default parameters (the full
black and full red curves, respectively). While the
unmatched generation well reproduces the 2-jet $\HT$, it increasingly
falls below the matched curve for $\HTjet(3)$ and $\HTjet(4)$. In real
data, this deficit could easily be interpreted as a sign for a missing
production mode, such as associated production with gluinos and
squarks, or gluino pair production. 

\FIGURE[t]{
\epsfig{file=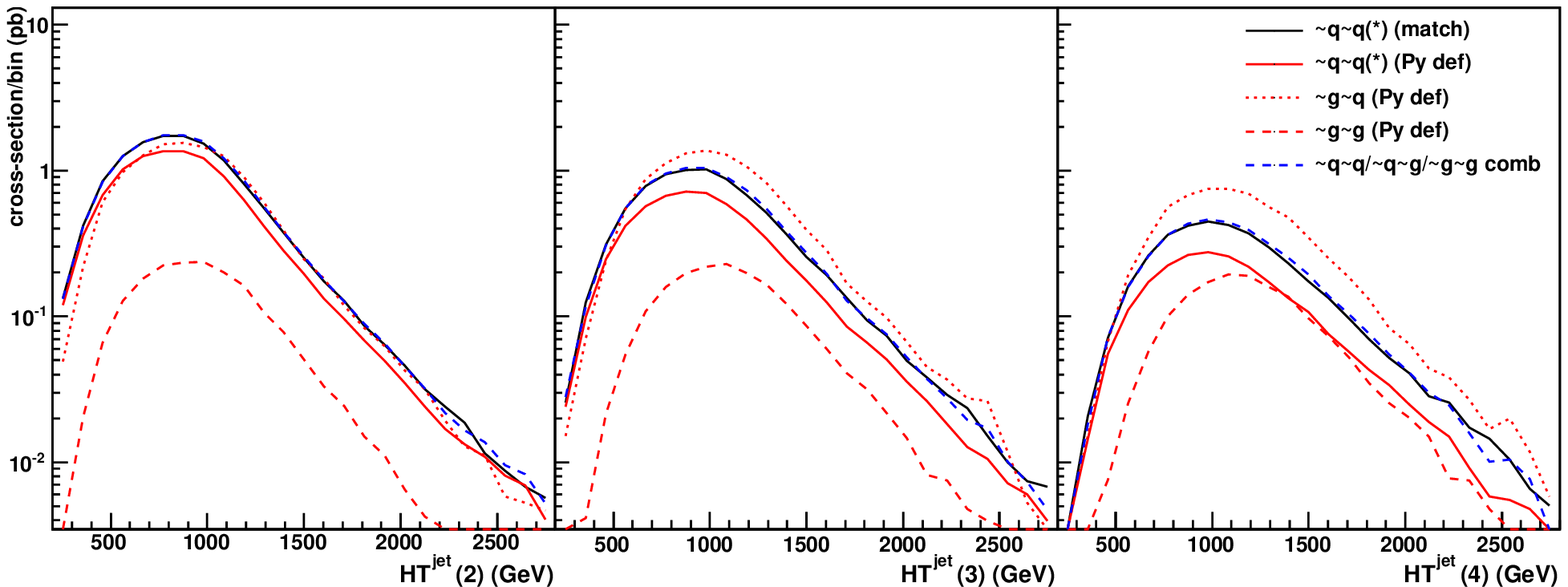,width=15cm}
\caption{$H_T(n)$ with $n=2,3,4$ for $\tilde q \tilde q^{(*)}$ 
production in Scenario III, with and without matching (full black and
full red curves), and for unmatched associated $\tilde g\tilde q$ and
$\tilde g\tilde g$ production with $m_{\tilde g}=700$ GeV(dotted and broken red curves). Also
shown is the result of adding the unmatched $\tilde q \tilde q^{(*)}$
curve and 0.25 $\times$ the $\tilde g\tilde q$ curve, which perfectly
mimicks the matched curve. All simulations are done with \pythia\
default $Q^2$-ordered showers.}
\label{fig:seven}
}

Also shown in Fig.~\ref{fig:seven} are the curves for associated gluino
production (dotted) and gluino pair production (broken red), with
gluinos at 700 GeV, decaying to the squarks emitting an additional
jet. While the cross section for the associated $\tilde g\tilde q$
production process is too large to be accommodated by the ``data''
(i.e.\ the matched squark production curve), there are many ways in
which this cross section estimate could be wrong; if, for example, the decay
modes are different for squarks and gluinos, or if the particles seen
are not from the MSSM but from some other realization of new
physics. We can therefore view the normalization of the different
curves as free parameters, and look at how the curves could be
combined to generate a fit to the ``data'' curve, i.e.\ the matched
$\tilde q\tilde q^{(*)}$ curve.

It turnes out that for these choices of masses, the data can be
very well accommodated by adding the gluino-squark associated production,
with the cross section reduced to 25\% of the nominal cross section, to
the unmatched squark production curve, as shown by the broken
blue line in the figure.

Possible cross checks to avoid this type of false
conclusions could be looking at detailed kinematic properties of the
jets, or looking at different decay signatures. However, this example still
illustrates the importance of using matching also for BSM signals, in
particular when looking at observables where extra jets from QCD
radiation may play an important role.

\subsection{Example 2: Degenerate spectrum}\label{ex2}
\label{sec:degenerate}

Another case where jet matching is of considerable importance is when
the mass splitting between the produced QCD particle and the LSP is
small, as is the case in Scenario IV. There, 600 GeV gluinos decay
through off-shell squarks to a neutral 550 GeV LSP and two quarks. The
main problem with searches in this scenario is that no large missing
$\ET$ is produced in the decay, since all decay products are soft. It
doesn't help if the gluinos are produced at a large invariant mass,
since they will then decay back-to-back, and the missing energy due to
the boost of the individual gluinos is canceled between the two
LSP's. While this type of scenario is not possible in typical
unification scenarios (e.g.~mSUGRA or mGMSB), a more model-independent
approach necessitates taking them into account
\cite{Alwall:2008va,Alwall:2008ve}.

The only way to get a large missing energy in this scenario is through
a transverse boost of the gluino pair center-of-mass system. This
happens when the gluino pair system recoils against hard initial state
radiation. In this case, a proper jet matching between parton showers
and matrix element QCD emission is crucial to well describe the $\MET$
distribution as well as the resulting jet structure. This is
particularly important since a $\MET$ of a certain magnitude as well
as several hard jets is typically needed to pass the experiment
triggers, so that a misrepresentation of these quantities will lead to
very different trigger acceptences.

\FIGURE[t]{
\epsfig{file=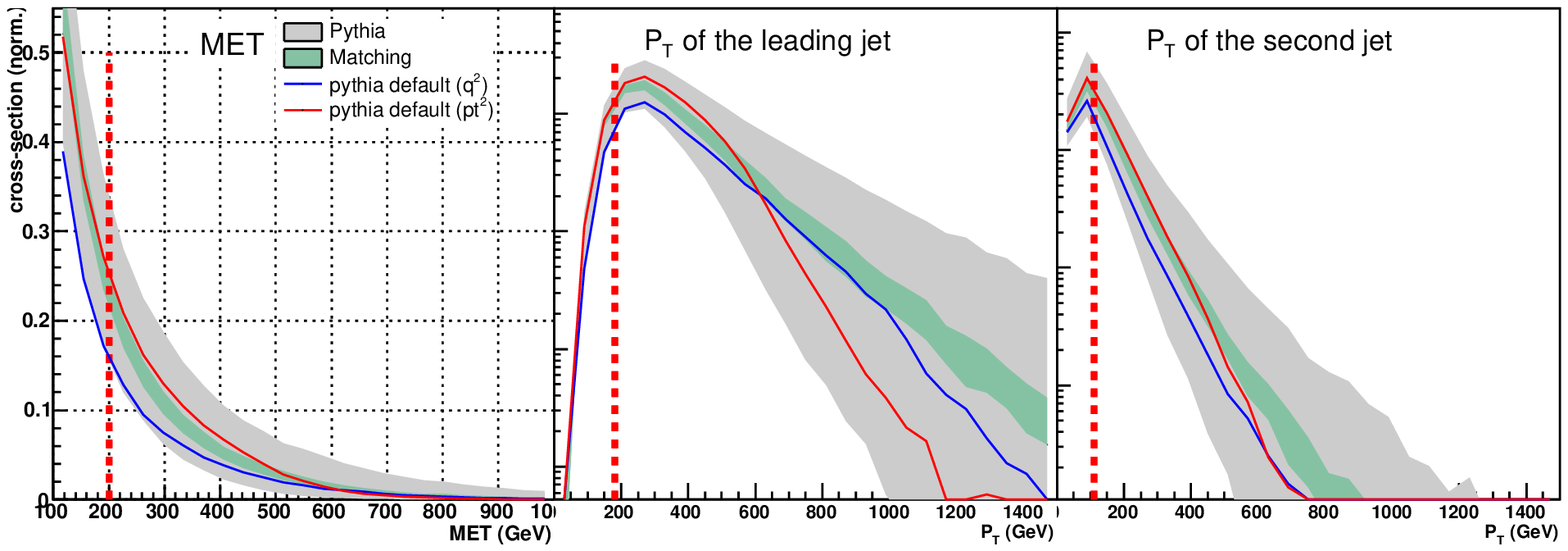,width=15cm}
\caption{$\MET$ and $\pt$ for the two hardest jets in
$\tilde g \tilde g$ production in Scenario IV. The grey band shows the
spread of unmatched \pythia\ predictions with varying shower
parameters, while the green band shows the corresponding matched
predictions. The full blue and red curves represents the default unmatched
\pythia\ $Q^2$- and $\pt$-ordered showers respectively.}
\label{fig:eight}
}

We show in Fig.~\ref{fig:eight} three representative quantities for
this case, the $\MET$ and the $\pt$ of the two hardest jets. As can be
seen from the figure, the virtuality-ordered \pythia\ default
distribution severely underestimates the missing $\ET$ as well as the
hardness of the leading jets, which mainly come from the initial state
radiation. The effect, with typical cuts jet-$\MET$ cuts like
$\MET>200$ GeV, $\pt^{\mathrm{jet}_1}>180$ GeV and $\pt^{\mathrm{jet}_2}>110$
GeV is a signal efficiency less then half that of the matched
production. The $\pt$-ordered shower rather overshoots the matrix
element curves, up to the factorization scale ($m_{\tilde g}$), where
it falls off rapidly. Once again, the lesson here is that in order to
get a description that is predictive and insensitive to the details of
the shower parameterization, it is necessary to use jet matching.


\section{Impact on BSM searches}
\label{sec:SPS1a}

We now consider the effects of an accurate simulation of QCD radiation
in the typical observables employed in the BSM inclusive searches:
high $\pt$ jets and high missing transverse energy.  In order to see
the effect of the matching on the sensitivity to shower after a
smearing of the signal, and in order to be as complete as possible, we
consider the production of $\tilde g \tilde g$, $\tilde g \tilde q$,
$\tilde q \tilde q^{(*)}$ and $\tilde t_{1,2}\tilde t_{1,2}^*$, with
$\tilde q$ defined as $\tilde u_{L,R}$, $\tilde d_{L,R}$, $\tilde
s_{L,R}$, $\tilde c_{L,R}$, $\tilde b_{1,2}$. The signal is 
produced in both the matched (2$\rightarrow$2, 3, 4) and unmatched
(2$\rightarrow$2) modes in the the SPS1a benchmark
scenario~\cite{Allanach:2002nj}. For the background, we consider the
most important processes leading to four hard jets and potentially
large missing transverse energy: $W^\pm\rightarrow l^\pm\nu+4$ jets,
$Z^0\rightarrow\nu\nu+4$ jets, $W^\pm\rightarrow\tau_{jet}\nu+3$ jets
and finally the inclusive $t\overline{t}$+0,1,2,3 jets. We do not
include QCD multijet production, since we have no means of realistically
performing simulations of the missing energy distribution, which is
due to decays of heavy quarks to neutrinos and jet mismeasurement in
the detector. We instead base our analysis on cuts similar to those
used in Refs.~\cite{CMS} and \cite{ATLAS}, and keep this contribution in
mind. All background simulations are done using jet matching. Many
comparisons have been done between matched and unmatched background
simulations, which are well known to differ by up to several orders of
magnitude for this type of multi-jet observables
\cite{Krauss:2004bs,Krauss:2005nu,Mangano:2006rw}. We therefore here
look at the effects of including matching only in the signal
simulation.

Detector simulation is performed using PGS 4 \cite{PGS4} with the
MidPoint cone algorithm with a minimum $\pt$ of 40 GeV and a radius of
0.5. To be conservative, we use only kinematic variables
associated to the jets and the missing transverse energy. The cuts
used are

\begin{itemize}
\item $N_\mathrm{jet}\ge 4$
\item $|\eta_1|<1.7$, $|\eta_{2,3,4...}|<3$
\item $\pt^{\mathrm{jet}_1}>$180 GeV, $\pt^{\mathrm{jet}_2}>110$ GeV,  $\pt^{\mathrm{jet}_{>2}}>50$ GeV
\item $\MET>150$ GeV
\item $\Delta\phi(\MET,\mathrm{jet}_1)>0.5$  and $\Delta\phi(\MET,\mathrm{jet}_2)>1$
\item $\sum_{i=2}^{4} \pt^{\mathrm{jet}_i}+\MET>600$ GeV \,.

\end{itemize}

\FIGURE[t]{
\epsfig{file=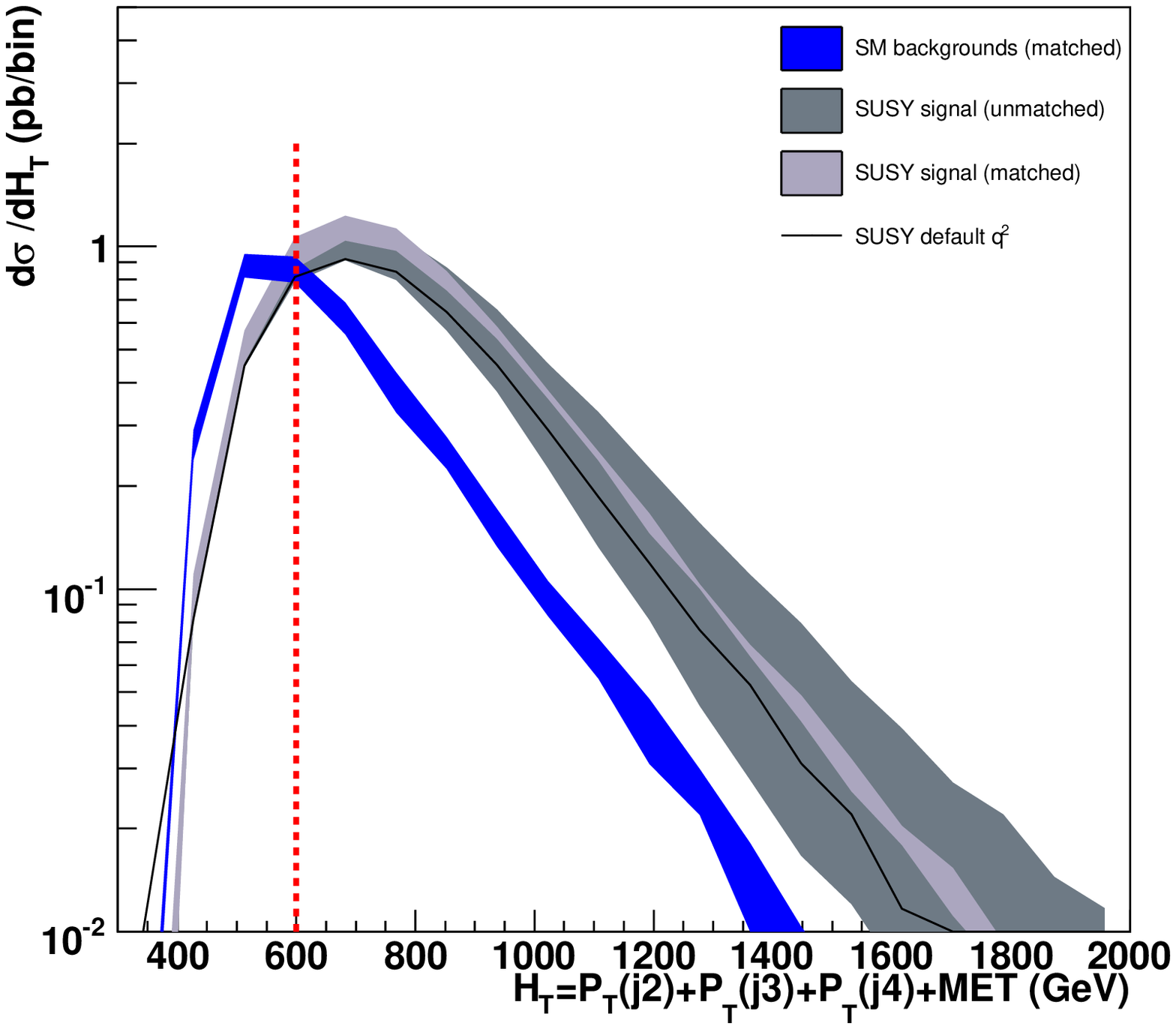,width=9cm}
\caption{$\HT$ for SUSY production in the SPS1a benchmark scenario,
compared to (matched) Standard Model backgrounds. The grey band shows the
spread of unmatched \pythia\ predictions with varying shower
parameters, while the green band shows the corresponding matched
predictions. $\HT$ definition and cuts are described in the text.}
\label{fig:bsm}
}

The SPS1a scenario is affected by several of the difficulties
described in Sec.~\ref{sec:anatomy}. The gluino has a mass higher
than, but close to, the squarks, and hence decays to a squark and a
soft jet, making the jet counting complicated. Since gluinos and
squarks are of similar mass, the QCD SUSY production includes
associated squark-gluino production, gluino pair production and squark
pair production (both $\tilde q\tilde q$ and $\tilde q\tilde g^*$
which are of similar cross section, and to a negligible degree
$\tilde q^*\tilde q^*$), in order of cross section. This means that
the \pythia\ shower cannot simultaneously describe all production modes,
as demonstrated in Fig.~\ref{fig:four}. Furthermore the separation of
the different production modes based on jet counts or jet kinematics
will be non-trivial.  The production cross section is however in this
scenario dominated by associated gluino-squark and gluino pair
production, where the default \pythia\ description is reasonably good,
and only undershoots the matched description by about 10-30\% for the
first couple of QCD radiation jets, so we expect the inclusive \pythia\
description to be reasonably close to the matched curve.

The result is illustrated in Fig.~\ref{fig:bsm},
which shows the $H_T=\sum_{i=2}^{4}\pt^{\mathrm{jet}_i}+\MET$ for inclusive
supersymmetric production of gluinos and squarks (including $\tilde{t}$ and $\tilde{b}$). 

The effect of the matching is as expected -- a significant reduction
in the sensitivity to parton shower parameters, and a shift of the
prediction as compared to the default virtuality-ordered \pythia\
shower by about 10-30\%. Even with the smearing due to the detector
simulation and more complex decays than the simplified scenarios used
in Sec.~\ref{sec:anatomy},
the ``power" $\pt$-ordered shower continues to overshoot the result
obtained with the matching whereas the ``wimpy'' virtuality-ordered
showers undershoot the matched curve.

Since we require four hard jets, the strongest impact is on the
squark-(anti)squark pair production, with an efficiency increase close
to 40\% when passing from the unmatched default virtuality-ordered
\pythia\ shower to the matched production. This happens for two reasons; as
described in Sec.~\ref{jetnumsec}, only two hard jets are produced by
the decay of the squarks, which means that two jets from QCD radiation
are needed. Second, as shown in Fig.~\ref{fig:four}, the difference
between the unmatched and matched radiation is particularly large for
squark production. On the other hand, the effect is much smaller for
gluino pair production (around $5\%$), which has to do mainly with the
large fraction of events with at least one top quark in the decay,
giving rise to multiple hard jets and hence a small sensitivity to the
matching.  The effect on the associated
gluino-squark production lies between these two extremes, with an
efficiency increase close to $10\%$. $\tilde{t}$ and $\tilde{b}$
production is dominated by $\tilde{t}_1\tilde{t}_1^*$ pairs (due
to a low $\tilde t_1$ mass, around 390 GeV), which decay to top quarks
and the LSP. For those, the effect of the matching is negligible.


\section{Conclusions}
\label{sec:conclusions}

Discovering new physics at the LHC will most probably be quite
challenging.  Apart from a few very clean but theoretically not very motivated
signatures, such as narrow resonances decaying into lepton pairs,
the vast majority of the theoretical constructions which are still
viable nowadays gives a rich and complex phenomenology that will be
hard to decipher.  For instance, many models that address the
hierarchy problem, predict new heavy colored states at the TeV scale
that can be easily produced at hadron colliders. To what extent the
decay products of such states will be identifiable on top of a large QCD
background, including vector boson(s) + jets, $t\bar t$ jets and
multi-jets, is currently subject of many theoretical and experimental
investigations.

In this work we have for the first time addressed the issue of
quantifying the effects of extra QCD radiation in the production of
heavy colored states employing inclusive multi-jet samples obtained by
matching multiparton matrix elements and parton showers.  Previous
work has either considered radiation in Standard Model $t\bar t$
events only~\cite{Mangano:2006rw} or in SUSY but at
parton level~\cite{Plehn:2005cq}.

Our results can be briefly summarized as follows.  First the extension
of matching techniques to beyond the Standard Model scenarios, such
as SUSY or UED, while posing no problems of principle, requires
dealing with new technical issues. For instance, the combination of
multi-parton samples for production of different resonances decaying
into jets
leads to
problems of double counting that need to be addressed. We have
proposed two working solutions, both of which have been implemented in
\mgme.

The main result of this work is that a matched matrix element plus
parton shower approach for heavy particle production is in general
much more accurate and predictive than a parton shower alone. We find
that, contrary to the ``common lore'' that showers alone provide a
good description of extra radiation when the produced states are
heavy, there are many cases where matrix element corrections are
indispensable. Not only are the
$\pt$ spectra of the extra jets in the parton shower approach
extremely sensitive to the shower starting scale and the shower algorithm,
while the matched simulations are not, but also that there is in general
no tuning of the shower that can simultaneously reproduce the matched
samples for all initial states.  We have presented 
examples where simulations based only on the parton shower could be
misleading. Another possibly relevant issue, that we leave to future
studies, is matching of radiation in the decays of the heavy states.
Though already possible in our current implementation, we have not
included it in this study mainly because it is computationally
expensive.  Next-to-leading order effects in decays have previously
been shown to be possibly important in precision studies, e.g.\ of the
spins of new particles \cite{Horsky:2008yi}. If found relevant,
also matching of radiation in decays could be included in future
simulations.

In conclusion, we recommend that matched samples should be used
not only for backgrounds but also for beyond the Standard Model
physics signals. Having the best available simulations could be
important not only in designing better and more solid strategies to
make discoveries but even more to identify the nature of the new
physics.
For precision measurements of new physics properties, as well as
reliably distinguish between different scenarios, fully differential
next-to-leading order simulations will most probably be needed. Given
the large number of theoretical possibilities still open, it is clear
that an automatic approach to NLO computations, and to their matching
with a parton shower, will be certainly welcome.

\section*{Acknowledgements}

The authors want to thank Gavin Salam, Maxim Perelstein and Tilman
Plehn for several interesting discussions. J.A.\ and F.M.\ would like to thank
the Aspen Center for Physics and the program ``LHC: Beyond the Standard
Model Signals in a QCD Environment'' where much of this work was
finalized. Big thanks also to Thomas Keutgen, Pavel Demin and Fabrice
Charlier for computer cluster support. J.A.\ was supported by the
Swedish Research Council.

\bibliography{database}

\providecommand{\href}[2]{#2}\begingroup\raggedright\begin{thebibliography}{10}

\bibitem{Sjostrand:2006za}
T.~{Sj\"ostrand}, S.~Mrenna, and P.~Skands, {\it {PYTHIA 6.4 physics and
  manual}},  {\em JHEP} {\bf 05} (2006) 026,
  [\href{http://xxx.lanl.gov/abs/hep-ph/0603175}{{\tt hep-ph/0603175}}].

\bibitem{Corcella:2000bw}
G.~Corcella {\em et~al.}, {\it Herwig 6: An event generator for hadron emission
  reactions with interfering gluons (including supersymmetric processes)},
  {\em JHEP} {\bf 01} (2001) 010,
  [\href{http://xxx.lanl.gov/abs/hep-ph/0011363}{{\tt hep-ph/0011363}}].

\bibitem{Gleisberg:2003xi}
T.~Gleisberg {\em et~al.}, {\it Sherpa 1.alpha, a proof-of-concept version},
  {\em JHEP} {\bf 02} (2004) 056,
  [\href{http://xxx.lanl.gov/abs/hep-ph/0311263}{{\tt hep-ph/0311263}}].

\bibitem{Pukhov:2004ca}
A.~Pukhov, {\it {CalcHEP 3.2: MSSM, structure functions, event generation,
  batchs, and generation of matrix elements for other packages}},
  \href{http://xxx.lanl.gov/abs/hep-ph/0412191}{{\tt hep-ph/0412191}}.

\bibitem{Boos:2004kh}
{\bf CompHEP} Collaboration, E.~Boos {\em et~al.}, {\it {CompHEP 4.4: Automatic
  computations from Lagrangians to events}},  {\em Nucl. Instrum. Meth.} {\bf
  A534} (2004) 250--259, [\href{http://xxx.lanl.gov/abs/hep-ph/0403113}{{\tt
  hep-ph/0403113}}].

\bibitem{Alwall:2007st}
J.~Alwall {\em et~al.}, {\it {MadGraph/MadEvent v4: The New Web Generation}},
  {\em JHEP} {\bf 09} (2007) 028,
  [\href{http://xxx.lanl.gov/abs/arXiv:0706.2334}{{\tt arXiv:0706.2334}}].

\bibitem{Kilian:2007gr}
W.~Kilian, T.~Ohl, and J.~Reuter, {\it {WHIZARD: Simulating Multi-Particle
  Processes at LHC and ILC}},
  \href{http://xxx.lanl.gov/abs/arXiv:0708.4233}{{\tt arXiv:0708.4233}}.

\bibitem{Catani:2001cc}
S.~Catani, F.~Krauss, R.~Kuhn, and B.~R. Webber, {\it Qcd matrix elements +
  parton showers},  {\em JHEP} {\bf 11} (2001) 063,
  [\href{http://xxx.lanl.gov/abs/hep-ph/0109231}{{\tt hep-ph/0109231}}].

\bibitem{Krauss:2002up}
F.~Krauss, {\it {Matrix elements and parton showers in hadronic interactions}},
   {\em JHEP} {\bf 08} (2002) 015,
  [\href{http://xxx.lanl.gov/abs/hep-ph/0205283}{{\tt hep-ph/0205283}}].

\bibitem{Lonnblad:2001iq}
L.~{L\"onnblad}, {\it {Correcting the colour-dipole cascade model with fixed
  order matrix elements}},  {\em JHEP} {\bf 05} (2002) 046,
  [\href{http://xxx.lanl.gov/abs/hep-ph/0112284}{{\tt hep-ph/0112284}}].

\bibitem{MLM}
M.~L. Mangano, ``Merging multijet matrix elements and shower evolution in
  hadronic collisions.'' Available at
  {\footnotesize\texttt{http://cern.ch/$\sim$mlm/talks/lund-alpgen.pdf}}, 2004.

\bibitem{Mangano:2006rw}
M.~L. Mangano, M.~Moretti, F.~Piccinini, and M.~Treccani, {\it {Matching matrix
  elements and shower evolution for top- quark production in hadronic
  collisions}},  {\em JHEP} {\bf 01} (2007) 013,
  [\href{http://xxx.lanl.gov/abs/hep-ph/0611129}{{\tt hep-ph/0611129}}].

\bibitem{Mrenna:2003if}
S.~Mrenna and P.~Richardson, {\it {Matching matrix elements and parton showers
  with HERWIG and PYTHIA}},  {\em JHEP} {\bf 05} (2004) 040,
  [\href{http://xxx.lanl.gov/abs/hep-ph/0312274}{{\tt hep-ph/0312274}}].

\bibitem{Alwall:2007fs}
J.~Alwall {\em et~al.}, {\it Comparative study of various algorithms for the
  merging of parton showers and matrix elements in hadronic collisions},  {\em
  Eur. Phys. J.} {\bf C53} (2008) 473--500,
  [\href{http://xxx.lanl.gov/abs/arXiv:0706.2569}{{\tt arXiv:0706.2569}}].

\bibitem{Krauss:2004bs}
F.~Krauss, A.~Sch{\"a}licke, S.~Schumann, and G.~Soff, {\it Simulating w / z +
  jets production at the tevatron},  {\em Phys. Rev.} {\bf D70} (2004) 114009,
  [\href{http://xxx.lanl.gov/abs/hep-ph/0409106}{{\tt hep-ph/0409106}}].

\bibitem{Frixione:2003ei}
S.~Frixione, P.~Nason, and B.~R. Webber, {\it Matching nlo qcd and parton
  showers in heavy flavour production},  {\em JHEP} {\bf 08} (2003) 007,
  [\href{http://xxx.lanl.gov/abs/hep-ph/0305252}{{\tt hep-ph/0305252}}].

\bibitem{Stelzer:1994ta}
T.~Stelzer and W.~F. Long, {\it Automatic generation of tree level helicity
  amplitudes},  {\em Comput. Phys. Commun.} {\bf 81} (1994) 357--371,
  [\href{http://xxx.lanl.gov/abs/hep-ph/9401258}{{\tt hep-ph/9401258}}].

\bibitem{Maltoni:2002qb}
F.~Maltoni and T.~Stelzer, {\it {MadEvent: Automatic event generation with
  MadGraph}},  {\em JHEP} {\bf 02} (2003) 027,
  [\href{http://xxx.lanl.gov/abs/hep-ph/0208156}{{\tt hep-ph/0208156}}].

\bibitem{Hagiwara:2005wg}
K.~Hagiwara {\em et~al.}, {\it {Supersymmetry simulations with off-shell
  effects for LHC and ILC}},  {\em Phys. Rev.} {\bf D73} (2006) 055005,
  [\href{http://xxx.lanl.gov/abs/hep-ph/0512260}{{\tt hep-ph/0512260}}].

\bibitem{Cho:2006sx}
G.~C. Cho {\em et~al.}, {\it {Weak boson fusion production of supersymmetric
  particles at the LHC}},  {\em Phys. Rev.} {\bf D73} (2006) 054002,
  [\href{http://xxx.lanl.gov/abs/hep-ph/0601063}{{\tt hep-ph/0601063}}].

\bibitem{Mangano:2002ea}
M.~L. Mangano, M.~Moretti, F.~Piccinini, R.~Pittau, and A.~D. Polosa, {\it
  Alpgen, a generator for hard multiparton processes in hadronic collisions},
  {\em JHEP} {\bf 07} (2003) 001,
  [\href{http://xxx.lanl.gov/abs/hep-ph/0206293}{{\tt hep-ph/0206293}}].

\bibitem{Catani:1993hr}
S.~Catani, Y.~L. Dokshitzer, M.~H. Seymour, and B.~R. Webber, {\it
  Longitudinally invariant k(t) clustering algorithms for hadron hadron
  collisions},  {\em Nucl. Phys.} {\bf B406} (1993) 187--224.

\bibitem{Tait:1999cf}
T.~M.~P. Tait, {\it {The t W- mode of single top production}},  {\em Phys.
  Rev.} {\bf D61} (2000) 034001,
  [\href{http://xxx.lanl.gov/abs/hep-ph/9909352}{{\tt hep-ph/9909352}}].

\bibitem{Frixione:2008yi}
S.~Frixione, E.~Laenen, P.~Motylinski, B.~R. Webber, and C.~D. White, {\it
  {Single-top hadroproduction in association with a W boson}},  {\em JHEP} {\bf
  07} (2008) 029, [\href{http://xxx.lanl.gov/abs/arXiv:0805.3067}{{\tt
  arXiv:0805.3067}}].

\bibitem{Plehn:2005cq}
T.~Plehn, D.~Rainwater, and P.~Skands, {\it {Squark and gluino production with
  jets}},  {\em Phys. Lett.} {\bf B645} (2007) 217--221,
  [\href{http://xxx.lanl.gov/abs/hep-ph/0510144}{{\tt hep-ph/0510144}}].

\bibitem{Odagiri:1998ep}
K.~Odagiri, {\it {Color connection structure of (supersymmetric) {QCD} ($2\to
  2$) processes}},  {\em JHEP} {\bf 10} (1998) 006,
  [\href{http://xxx.lanl.gov/abs/hep-ph/9806531}{{\tt hep-ph/9806531}}].

\bibitem{TuneA}
R.~Field, {\it {Pythia Tunes for Tevatron Run 2}}, . See
  {\footnotesize\texttt{http://www.phys.ufl.edu/$\sim$rfield/cdf/}}.

\bibitem{Barr:2007hy}
A.~J. Barr, B.~Gripaios, and C.~G. Lester, {\it {Weighing Wimps with Kinks at
  Colliders: Invisible Particle Mass Measurements from Endpoints}},  {\em JHEP}
  {\bf 02} (2008) 014, [\href{http://xxx.lanl.gov/abs/arXiv:0711.4008}{{\tt
  arXiv:0711.4008}}].

\bibitem{Cho:2007dh}
W.~S. Cho, K.~Choi, Y.~G. Kim, and C.~B. Park, {\it {Measuring superparticle
  masses at hadron collider using the transverse mass kink}},  {\em JHEP} {\bf
  02} (2008) 035, [\href{http://xxx.lanl.gov/abs/arXiv:0711.4526}{{\tt
  arXiv:0711.4526}}].

\bibitem{Lafaye:2004cn}
R.~Lafaye, T.~Plehn, and D.~Zerwas, {\it {SFITTER: SUSY parameter analysis at
  LHC and LC}},  \href{http://xxx.lanl.gov/abs/hep-ph/0404282}{{\tt
  hep-ph/0404282}}.

\bibitem{Lester:2005je}
C.~G. Lester, M.~A. Parker, and M.~J. White, {\it {Determining SUSY model
  parameters and masses at the LHC using cross-sections, kinematic edges and
  other observables}},  {\em JHEP} {\bf 01} (2006) 080,
  [\href{http://xxx.lanl.gov/abs/hep-ph/0508143}{{\tt hep-ph/0508143}}].

\bibitem{Miller:2005zp}
D.~J. Miller, P.~Osland, and A.~R. Raklev, {\it {Invariant mass distributions
  in cascade decays}},  {\em JHEP} {\bf 03} (2006) 034,
  [\href{http://xxx.lanl.gov/abs/hep-ph/0510356}{{\tt hep-ph/0510356}}].

\bibitem{Horsky:2008yi}
R.~Horsky, M.~Kramer, A.~Muck, and P.~M. Zerwas, {\it {Squark Cascade Decays to
  Charginos/Neutralinos: Gluon Radiation}},  {\em Phys. Rev.} {\bf D78} (2008)
  035004, [\href{http://xxx.lanl.gov/abs/arXiv:0803.2603}{{\tt
  arXiv:0803.2603}}].

\bibitem{Allanach:2002nj}
B.~C. Allanach {\em et~al.}, {\it {The Snowmass points and slopes: Benchmarks
  for SUSY searches}},  \href{http://xxx.lanl.gov/abs/hep-ph/0202233}{{\tt
  hep-ph/0202233}}.

\bibitem{Salam:2007xv}
G.~P. Salam and G.~Soyez, {\it {A practical Seedless Infrared-Safe Cone jet
  algorithm}},  {\em JHEP} {\bf 05} (2007) 086,
  [\href{http://xxx.lanl.gov/abs/arXiv:0704.0292}{{\tt arXiv:0704.0292}}].

\bibitem{Alwall:2008va}
J.~Alwall, M.-P. Le, M.~Lisanti, and J.~G. Wacker, {\it {Model-Independent Jets
  plus Missing Energy Searches}},
  \href{http://xxx.lanl.gov/abs/arXiv:0809.3264}{{\tt arXiv:0809.3264}}.

\bibitem{Alwall:2008ve}
J.~Alwall, M.-P. Le, M.~Lisanti, and J.~G. Wacker, {\it {Searching for Gluinos
  at the Tevatron}},  {\em Phys. Rev. Lett.} {\bf B666} (2008) 34--37,
  [\href{http://xxx.lanl.gov/abs/arXiv:0803.0019}{{\tt arXiv:0803.0019}}].

\bibitem{CMS}
{\bf CMS} Collaboration, G.~L. Bayatian {\em et~al.}, {\it {CMS technical
  design report, volume II: Physics performance}},  {\em J. Phys.} {\bf G34}
  (2007) 995--1579.

\bibitem{ATLAS}
{ATLAS Collaboration}, {\it {ATLAS detector and physics performance. Technical
  design report. Vol. 2}}, . CERN-LHCC-99-15.

\bibitem{Krauss:2005nu}
F.~Krauss, A.~Schalicke, S.~Schumann, and G.~Soff, {\it {Simulating W / Z +
  jets production at the CERN LHC}},  {\em Phys. Rev.} {\bf D72} (2005) 054017,
  [\href{http://xxx.lanl.gov/abs/hep-ph/0503280}{{\tt hep-ph/0503280}}].

\bibitem{PGS4}
J.~Conway {\em et~al.}, {\it {PGS: Pretty Good Simulation of high energy
  collisions}}, . See
  {\footnotesize\texttt{http://www.physics.ucdavis.edu/$\sim$conway/research/s%
oftware/pgs/pgs4-general.htm}}.

\end{thebibliography}\endgroup

\end{document}